\newif\ifconfver
\confvertrue        %declaring conference version u

\ifconfver
\documentclass[10pt,twocolumn,twoside]{IEEEtran}
\else
\documentclass[11pt,draftcls,onecolumn]{IEEEtran}
\fi

\usepackage{tikz }
\usepackage{quantikz}
\usepackage{makecell}
\usepackage{bbding}
\usepackage{color,graphicx}
\usepackage{float}

\usepackage{multirow,amsmath,epsfig,amsfonts,amssymb,psfig,graphics,psfrag,theorem,calc,url,bm,cite}
\usepackage{stfloats,hyperref}
\usepackage{algorithm}     
\usepackage{algpseudocode}
\usepackage{booktabs}  %  用來畫table
\usepackage{pifont} % 用來ablation的打勾跟打叉符號
\usepackage{diagbox} 
\usepackage{hhline}
\usepackage{subfigure}
\usepackage{comment}     
\usepackage{threeparttable}
\usepackage{caption}
\usepackage{fancyhdr}

\usepackage{mathtools}

\makeatletter
\def\multilimits@{\bgroup
\Let@
\restore@math@cr
\default@tag
\baselineskip\fontdimen10 \scriptfont\tw@
\advance\baselineskip\fontdimen12 \scriptfont\tw@
\lineskip\thr@@\fontdimen8 \scriptfont\thr@@
\lineskiplimit\lineskip
\vbox\bgroup\ialign\bgroup\hfil$\m@th\scriptstyle{##}$\hfil\crcr}
\def\Sb{_\multilimits@}
\def\endSb{\crcr\egroup\egroup\egroup}
\makeatother

{\begin{list}{}%
{\setlength{\rightmargin}{0pc}%
\setlength{\leftmargin}{1pc} }} %
{\end{list}}

\newlength{\twidth}
\ifconfver
\else
\fi
\definecolor{orange}{RGB}{255,107,0}

\newtheorem{theorem}{Theorem}

\theorembodyfont{\rmfamily}

{\begin{list}{}{
			\settowidth{\labelwidth}{\mbox{\textnormal{#1}}}%
			\setlength{\leftmargin}{\labelwidth+\labelsep}}}%
	{\end{list}}

\newcommand\bA{\ensuremath{{\bm A}}}
\newcommand\bB{\ensuremath{{\bm B}}}
\newcommand\bC{\ensuremath{{\bm C}}}

\newcommand\bF{\ensuremath{{\bm F}}}

\newcommand\bI{\ensuremath{{\bm I}}}

\newcommand\bS{\ensuremath{{\bm S}}}

\newcommand\bZ{\ensuremath{{\bm Z}}}

\newcommand\bh{\ensuremath{{\bm h}}}

\newcommand\bs{\ensuremath{{\bm s}}}

\newcommand\bx{\ensuremath{{\bm x}}}
\newcommand\by{\ensuremath{{\bm y}}}
\newcommand\bz{\ensuremath{{\bm z}}}

\definecolor{orange}{RGB}{255,107,0}

\ifconfver
\author{Po-Wei Tang,~\IEEEmembership{Student Member,~IEEE}, Chia-Hsiang Lin,~\IEEEmembership{Senior Member,~IEEE}, \\
Jian-Kai Huang,~\IEEEmembership{Student Member,~IEEE}, and Alfredo R. Huete}
\title{A Quantum-Empowered SPEI Drought Forecasting Algorithm Using Spatially-Aware Mamba Network
\thanks{This study was supported in part by the Emerging Young Scholar Program (namely, the 2030 Cross-Generation Young Scholars Program) of the National Science and Technology Council (NSTC), Taiwan, under Grant NSTC 113-2628-E-006-003; in part by the Ph.D. Students Study Abroad Program of NSTC under Grant NSTC 113-2917-I-006-011; and in part by the University Academic Alliance in Taiwan
(UAAT) Program of the Ministry of Education (MoE), Taiwan, under Grant H113-A-235.
We thank the National Center for Theoretical Sciences (NCTS) and the National Center for High-performance Computing (NCHC) for providing the computing resources.}
\thanks{\textit{(Corresponding author: Chia-Hsiang Lin.)}}
\thanks{Po-Wei Tang is with the Institute of Computer and Communication Engineering, Department of Electrical Engineering, National Cheng Kung University, Tainan 70101, Taiwan (R.O.C.) 
(e-mail:  q38091526@gs.ncku.edu.tw).}
\thanks{Chia-Hsiang Lin. Lin is with the Department of Electrical Engineering, and with the Miin Wu School of Computing, National Cheng Kung University, Tainan 70101, Taiwan (R.O.C.) 
(e-mail: chiahsiang.steven.lin@gmail.com).}
\thanks{Jian-Kai Huang is with the Institute of Computer and Communication Engineering, Department of Electrical Engineering, National Cheng Kung University, Tainan 70101, Taiwan (R.O.C.) 
(e-mail:  q36121147@gs.ncku.edu.tw).}
\thanks{Alfredo R. Huete is with the School of Life Sciences, University of Technology Sydney, Sydney, NSW 2007, Australia 
(e-mail: Alfredo.Huete@uts.edu.au).}	
}
\fancypagestyle{IEEEtitlepagestyle}{
	\fancyhf{} % 清除預設的頁眉和頁腳
	\fancyhead[LE,RO]{\thepage} % 保留頁眉的頁碼樣式
	\fancyfoot[C]{\scriptsize © 2025 IEEE. Personal use of this material is permitted. Permission from IEEE must be obtained for all other uses, in any current or future media, including reprinting/republishing this material for advertising or promotional purposes, creating new collective works, for resale or redistribution to servers or lists, or reuse of any copyrighted component of this work in other works. This is the accepted version of the manuscript. The final, published version is available at IEEE Xplore via https://doi.org/10.1109/TGRS.2025.3583578.}
	
}
\else
\fi
\begin{document}
\bibliographystyle{IEEEtran}
\maketitle
\ifconfver \else %\vspace{-0.5cm}
\fi
\begin{abstract}
Due to the intensifying impacts of extreme climate changes, drought forecasting (DF), which aims to predict droughts from historical meteorological data, has become increasingly critical for monitoring and managing water resources. 
%
% Though drought conditions often exhibit spatial climatic coherence among neighboring regions, benchmark deep learning-based DF methods overlook this fact and predict the conditions on a region-by-region basis.
%
Despite the spatial coherence of drought conditions, benchmark deep learning-based DF models predict each region independently while ignoring the neighboring spatial information.
Using the Standardized Precipitation Evapotranspiration Index (SPEI), we designed and trained a novel and transformative spatially-aware DF neural network, which effectively captures local interactions among neighboring regions, resulting in enhanced spatial coherence and prediction accuracy.
As DF also requires sophisticated temporal analysis, the Mamba network, recognized as the most accurate and efficient existing time-sequence modeling, was adopted to extract temporal features from short-term time frames.
We also adopted quantum neural networks (QNN) to entangle the spatial features of different time instances, leading to refined spatiotemporal features of seven different meteorological variables for effectively identifying short-term climate fluctuations.
In the last stage of our proposed SPEI-driven quantum spatially-aware Mamba network (SQUARE-Mamba), the extracted spatiotemporal features of seven different meteorological variables were fused to achieve more accurate DF.
Validation experiments across El Niño, La Niña, and normal years demonstrated the superiority of the proposed SQUARE-Mamba, remarkably achieving an average improvement of more than 9.8\% in the coefficient of determination index (R$^2$) compared to baseline methods, thereby illustrating the promising roles of the temporal quantum entanglement and Mamba temporal analysis to achieve more accurate DF. 
Notably, the integration of QNN further upgrades the naive Mamba baseline by over 2.7\% in R$^2$ on average, highlighting the model’s sensitivity to transient climate variations.
Source codes: \url{https://github.com/IHCLab/SQUARE-Mamba}.

%---------
\bfseries{\em Index Terms---}Drought Forecasting, 
Quantum Computing, 
Mamba Deep Learning,
Climatic Research Unit (CRU) Data,
Standardized Precipitation Evapotranspiration Index (SPEI),
Sustainable Development Goals (SDGs).
%---------
\end{abstract}
\ifconfver \else \vspace{-0.0cm}\fi
\ifconfver \else \vspace{-0.5cm}\fi
%
%ken does some tricks with line spacing: end here
\ifconfver \else  \fi

\section{Introduction} \label{sec:introduction}
Extreme climate changes have altered precipitation patterns, leading to more frequent and severe drought occurrences.
Consequently, many regions worldwide have experienced significant agricultural losses and water supply challenges \cite{wilhite2016drought,mao2015climate}.
Nevertheless, these widespread impacts cannot be attributed solely to climatic changes.
Human activities, including urbanization, deforestation, and agricultural expansion, also play a critical role in modifying hydrological processes and land surfaces, thereby influencing the frequency and severity of droughts in human-altered regions such as America, Brazil, China, Spain, and Australia \cite{van2016drought}.
Recent literature indicates that urban areas are expected to accommodate an additional 2 billion people by 2030, significantly increasing pressure on global water resources \cite{ImmerzealScience2010,taylor2013ground}.
Moreover, more than 27\% of the world’s largest cities are projected to face severe water shortages, potentially exhausting their available water supplies by 2050 \cite{florke2018water}.
Beyond water scarcity, drought presents critical challenges for agriculture, industry, public health, and ecosystems \cite{giovannini2016drought,Xiang2023,chang2016climate,bond2008impacts}.

Given these widespread impacts, accurate drought forecasting (DF) has become an urgent task, aligning with the objectives of the United Nations Sustainable Development Goals (SDGs) 6, 11, 12, 13, and 15 \cite{zhang2019urban}.
In light of this urgency, it is essential to understand the different types of droughts, as each has distinct causes and impacts.
Specifically, droughts can be categorized into four main types: meteorological drought, caused by a prolonged scarcity of precipitation; agricultural drought, resulting from insufficient soil moisture, which adversely affects crop yields; hydrological drought, characterized by a significant reduction in streamflow and groundwater levels; and socioeconomic drought, which occurs when water scarcity leads to severe economic and social impacts \cite{salvia2021added}.
To better understand drought conditions and mitigate their adverse effects, researchers have developed various drought indices to assess drought severity.

For example, McKee \textit{et al.} \cite{mckee1993relationship} proposed the Standardized Precipitation Index (SPI) to assess wet and dry conditions based on precipitation data.
The SPI measures the deviation of observed precipitation from the long-term historical mean over different time scales, such as 1, 3, 6, or 12 months.
First, precipitation data is fitted to a gamma distribution to model the precipitation probability distribution, after which the cumulative probability of the precipitation values is determined.
Then, the cumulative probability is transformed into a standard normal distribution (with a mean of zero and a standard deviation of one) using an inverse standard normal function.
Finally, the resulting value represents the SPI for the specific precipitation data point.
Positive SPI values indicate wetter-than-average conditions, while negative values reflect drier-than-average conditions. 
Larger absolute values correspond to more severe deviations from the long-term historical mean.

By incorporating potential evapotranspiration (PET), Vicente-Serrano \textit{et al.} \cite{vicente2010multiscalar,vicente2012performance} proposed an enhanced version of the SPI, known as the Standardized Precipitation Evapotranspiration Index (SPEI).
Unlike the SPI, the SPEI accounts for the influence of temperature on drought conditions, providing a more comprehensive assessment.
In addition to developing drought indices based on meteorological data,  Huete \cite{HUETE1988295} proposed the Soil-Adjusted Vegetation Index (SAVI), an optical greenness measure of vegetation health, derived from remote sensing data, which is designed to minimize the influence of soil brightness on vegetation reflectance measurements.
SAVI serves as an indirect indicator of drought impacts on vegetation, as drought conditions typically induce vegetation stress, affecting NIR and red reflectance.
Given vegetation’s high sensitivity to drought, SAVI provides valuable insights into ecosystem responses under drought conditions.

Considering temperature variability, the SPEI is a practical approach for evaluating drought conditions.
Hence, multiple DF deep learning models based on SPEI have been developed, including the long short-term memory network (LSTM) \cite{hochreiter1997long}, the transformer-based time series prediction model (Informer) \cite{zhou2021informer}, and the autoregressive integrated moving average model (ARIMA) \cite{box2015time}.
For example, Dikshit \textit{et al.} \cite{DIKSHIT2021111979} proposed an improved SPEI-based DF approach using the LSTM neural network (NN).
The LSTM model predicts SPEI values based on hydrometeorological variables, including precipitation, maximum temperature, minimum temperature, mean temperature, vapor pressure, cloud cover, and potential evapotranspiration.
LSTM is an advanced variant of the recurrent NN (RNN) \cite{medsker1999recurrent}, which is designed to capture both long-term and short-term memory. 
It addresses the issues of gradient exploding/vanishing by using the customized forget gate, input gate, and output gate.
By extending the fully connected LSTM (FC-LSTM) to have convolutional structures, Shi \textit{et al.} \cite{ConvLSTM} proposed the convolutional LSTM (ConvLSTM) for the precipitation nowcasting problem.
Similar to LSTM, a variant of RNN using the gated recurrent unit (GRU) \cite{chung2014empirical} is proposed for predicting time series data.
In GRU, three gates in LSTM are replaced by two gates: the reset gate and the update gate.
Specifically, the reset gate controls how much of the previous state should be forgotten, while the update gate regulates the retention of information from the previous hidden state, helping the RNN capture long-term information \cite{cho2014learning}.

Based on the recently developed transformer model, Shang \textit{et al.} \cite{shang2023application} proposed the SPEI-based DF method using the Informer framework. 
The Informer maintains the advantages of the transformer structure, which is adept at capturing long-term dependencies and demonstrates strong potential for forecasting time-series data.
To reduce computational complexity and memory usage compared to the original transformer model, Informer utilizes the ProbSparse self-attention mechanism \cite{dong2022prediction} and an attention-distilling strategy.
In contrast, Wang \textit{et al.} \cite{simplemamba} introduced Simple-Mamba (S-Mamba), a Mamba-based model for time series forecasting, achieving state-of-the-art performance while significantly reducing computational burden compared to traditional Transformers \cite{dosovitskiy2021an}.
From another perspective, Hasan \textit{et al.} \cite{hasan2023spi} proposed another SPEI-based DF model based on the ARIMA.
It consists of three main components, including the autoregressive (AR) part, the integrated (I) part, and the moving average (MA) part.
The AR part makes predictions based on previous $p$-lag observations.
The I part transforms the data into a stationary series through $d$-order differencing operations.
The MA part adjusts the forecasting results based on the past $q$-lag forecasting errors \cite{hyndman2018forecasting}.

Beyond traditional forecasting methods, quantum technology has created new opportunities for algorithm development.
As quantum technology advances rapidly, it has become a cutting-edge scientific research area.
In 2022, the Nobel Prize in Physics was awarded to scientists who made significant contributions to quantum entanglement studies, highlighting the growing recognition of quantum information science.
As a result, quantum technology has recently been applied to remote sensing areas, particularly in addressing challenging inverse problems such as hyperspectral satellite data restoration\cite{hyperqueen}, mangrove mapping\cite{lin2024qmm}, change detection\cite{lin2024qhcd}, multispectral unmixing\cite{lin2024prime}, and well-log interpretation\cite{liu2021quantum}. 
In these tasks, the quantum neural network (QNN) has proven its value and applicability, demonstrating the potential of quantum technology in handling complicated problems.

Alongside these developments, recent studies have investigated integrating quantum computing into classical models.
For example, Pasetto \textit{et al.} \cite{pasetto2024kernel} introduced the adiabatic quantum kitchen sinks (AQKS) approach, a kernel-based approximation method that uses parallel quantum annealing on the D-Wave system to predict chlorophyll concentration in aquatic environments.
For cloud detection, Miroszewski \textit{et al.} \cite{miroszewski2023detecting} proposed a hybrid support vector machine (SVM) with quantum kernels using Landsat-8 multispectral imagery.
For classification tasks, Fan \textit{et al.} \cite{fan2025hybrid} developed a hybrid quantum deep learning framework featuring a superpixel-based encoding strategy that notably reduces quantum resource usage. 
Besides, Sebastianelli \textit{et al.} \cite{sebastianelli2021} showed that integrating entanglement into quantum circuit design enhances multiclass classification on the EuroSAT dataset.
To improve classification accuracy, Sebastianelli \textit{et al.} \cite{sebastianelli2025} proposed the Quanv4EO framework, which uses a quanvolution method \cite{henderson2020} for preprocessing multi-dimensional EO data. 
Beyond classification, Lin \textit{et al.} \cite{hyperking} recently introduced hybrid quantum-classical networks using a generative adversarial network (GAN) structure for hyperspectral tensor completion, mixed noise removal, and blind source separation.
From another perspective, Otgonbaatar \textit{et al.} \cite{otgonbaatar2023} examined the challenges and potential of quantum learning techniques in satellite data, focusing on balancing computational workloads between high-performance computing (HPC) and quantum computing (QC).

Beyond remote sensing applications, QNN has also achieved success in prediction-like tasks, including solar irradiance forecasting\cite{Hong2024qsif}, financial prediction\cite{PAQUET2022116583}, weather forecasting\cite{Safari2021qwf}, transport forecasting \cite{qu2022temporal}, cloud workloads \cite{gupta2024multiple}, and energy demand \cite{habibi2025electrical}.
Overall, these studies highlight a growing trend of applying quantum computing to time-series forecasting tasks, with satisfactory prediction outcomes.
Besides, motivated by information theory \cite{lin2024qmm,lin2024qhcd}, we integrate QNNs not as a replacement but as a complementary module that enriches feature distillations through unitary transformations under a hybrid quantum-classical forecasting framework.
Thus, we integrate QNN and convolutional NN (CNN) models to enhance prediction accuracy in the final decision-making process.
Specifically, our quantum component offers a theoretical guarantee of quantum expressibility.
Although the Ising quantum gate is introduced to enhance qubit interactions, our QNN framework retains full expressibility, as mathematically guaranteed in Theorem \ref{theorem: FE}.
Unlike existing quantum approaches that primarily utilize QNNs for spatial feature extraction, such as in hyperspectral image restoration \cite{hyperqueen}, hyperspectral change detection \cite{lin2024qhcd}, mangrove mapping \cite{lin2024qmm}, and image classification \cite{fan2023hybrid}, our method uniquely employs QNNs to model temporal dependencies.
To the best of our knowledge, this is the first application of QNNs in time-series modeling on the DF problem, and the experimental results demonstrate that integrating QNNs is particularly effective.

To further enhance the efficiency of the classical component within our hybrid quantum-classical framework, we integrate the Mamba network\cite{gu2023mamba}, a newly developed time-series model recognized for its state-of-the-art performance across various applications, including time-series forecasting\cite{liang2024bi}, natural language processing\cite{yue2024biomamba}, and image analysis\cite{Li2024mambahsi}.
Unlike traditional transformer-based models\cite{vaswani2017attention}, which rely heavily on self-attention mechanisms\cite{tang2024transformer,Young2024tnnls} with quadratic computational complexity, the Mamba network introduces an innovative approach by employing a linear time complexity architecture, significantly reducing computation when processing long sequences.
One key feature of Mamba is its ability to dynamically adjust state space model (SSM) parameters based on the input data, enabling the model to retain or discard information throughout the selective mechanism (cf. Section \ref{sec:method}).
This adaptive mechanism effectively addresses challenges commonly encountered with discrete modalities, improving learning efficiency.
Furthermore, its efficiency does not come at the cost of performance.
Mamba has demonstrated superior results in language modeling, outperforming transformers of the same size and matching the performance of models twice as large in pretraining and downstream evaluations.
As a result, Mamba achieves nearly an order-of-magnitude improvement in inference speed compared to transformer-based models, making it particularly well-suited for forecasting tasks such as the DF problem.

Given Mamba's remarkable efficiency in processing long sequences, it enhances our framework's temporal modeling capabilities, which are essential for complicated forecasting tasks such as the DF problem.
This capability has become increasingly important due to the growing impact of extreme climate changes.
Thus, we propose an SPEI-driven quantum spatially-aware Mamba network (SQUARE-Mamba) to effectively solve the challenging DF problem.
The main contributions of this work are summarized as follows:

%%===================Framework==============%%
\begin{figure*}[t]
\centering
\includegraphics[width=1\textwidth]{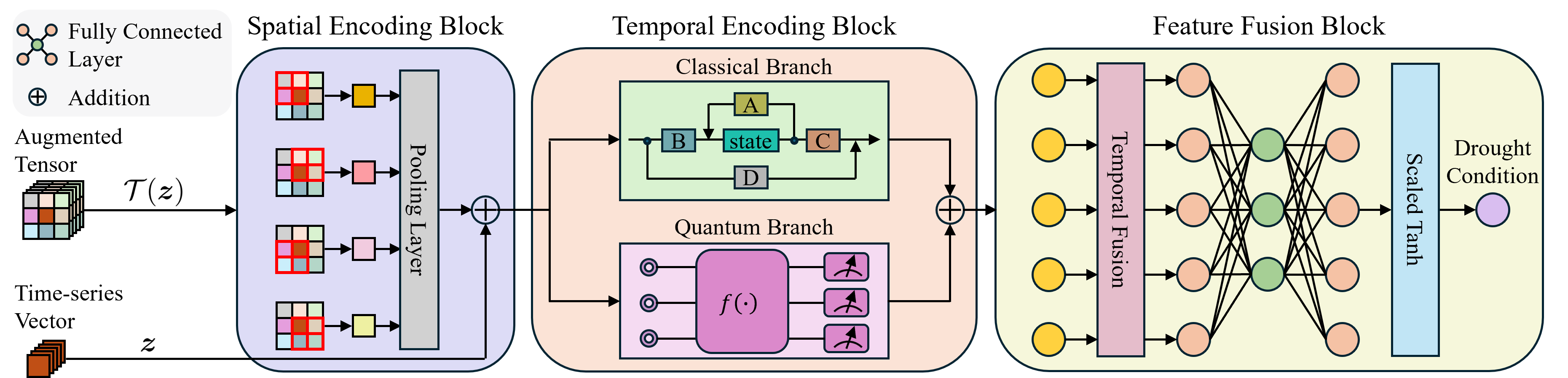}
%\vspace{-0.5cm}
\caption{Graphical illustration of the proposed SPEI-driven quantum spatially-aware Mamba network (SQUARE-Mamba).
It first incorporates neighboring spatial information through the spatial encoding block (SEB), which accepts the spatially augmented tensor $\mathcal{T}(\bz)$ as its input.
Next, the temporal encoding block (TEB) models temporal dependencies via a parallel CNN/QNN structure, in order to simultaneously incorporate the powerful Mamba temporal modeling and the quantum temporal entangling mechanisms.
Finally, the feature fusion block (FFB) integrates the extracted spatiotemporal features to forecast the drought prediction results.}
\label{fig:DFframework}
\end{figure*}

\begin{itemize}

\item 
To effectively capture the spatial continuity of meteorological data across adjacent regions, we have developed a spatial encoding block (SEB) to extract relevant spatial information from neighboring areas for each forecasting location.
SEB enables SQUARE-Mamba to distill critical spatial features, resulting in better spatial coherence and promising drought prediction accuracy.

\vspace{0.1cm}

\item 
In SQUARE-Mamba, a hierarchical structure is leveraged with a temporal encoding block (TEB) and a feature fusion block (FFB).
Within the TEB, SQUARE-Mamba elegantly integrates the novel QNN and Mamba models to enhance its temporal modeling capabilities.
Specifically, the input sequence for the TEB is partitioned into local groups, enabling the model to capture short-term dependencies and temporal variations across neighboring months.
This design improves the model robustness to short-term extreme climate fluctuations.
Empowered by the temporal quantum entanglement across different time instances for better capturing the short-term temporal dependencies, FFB aggregates the SEB/TEB-extracted global spatiotemporal features to generate the final forecasting results.

\vspace{0.1cm}

\item 
Our proposed novel SQUARE-Mamba leverages deep learning and quantum computing to achieve state-of-the-art DF performance, having a remarkable improvement of up to about $9.8\%$ in the R$^2$ index.
To ensure the effectiveness of SQUARE-Mamba, we evaluate its performance across diverse spatiotemporal scenarios.
From a spatial perspective, we test six datasets across three distinct climate regions (i.e., wet, dry, and moderate areas) to assess SQUARE-Mamba’s generalization ability to different environmental conditions.
From a temporal perspective, we analyze SQUARE-Mamba’s performance under El Niño, La Niña, and normal years to evaluate its stability across varying climate patterns.
Extensive experiments demonstrate that SQUARE-Mamba consistently outperforms existing benchmark DF methods across diverse environmental regions and climate phases.
By integrating SEB and TEB, SQUARE-Mamba offers a practical and reliable solution to the challenging DF problem.

\item In particular, our work further leverages quantum AI to better capture severe short-term climate fluctuations, which existing models such as LSTM and Transformer often fail to predict accurately.
For example, in the Geehi dataset (cf. Figure \ref{fig:wet_2}), the ground-truth curve shows sharp climate changes from March to December 2015 during the most intense El Niño event since 1950.
Despite their strengths, LSTM and Transformer predictions drift significantly during this period, missing these rapid variations.
In contrast, our model provides more accurate forecasts, enabling earlier warnings and reducing potential losses.
\end{itemize}

The remainder of this paper is organized as follows.
In Section \ref{sec:method}, we provide a detailed description of the proposed SQUARE-Mamba model.
Section \ref{sec:experiment} presents and discusses extensive experimental results conducted across El Niño, La Niña, and normal years over a long time period.
Besides, we perform ablation studies to demonstrate the importance of incorporating spatial information, and to evaluate the effectiveness of multi-scale temporal features of quantum-empowered Mamba.
Finally, Section \ref{sec:conclusion} summarizes the conclusions and key insights of this study.

\section{The Proposed SQUARE-Mamba Method for Drought Forecasting} \label{sec:method}

To address the limitations of existing benchmark DF models, which tend to overlook the spatial coherence of climatic patterns, we propose a radically new SPEI-driven quantum spatially-aware Mamba neural network (i.e., SQUARE-Mamba network).
As shown in Figure \ref{fig:DFframework}, SQUARE-Mamba comprises three key modules, including the spatial encoding block (SEB), temporal encoding block (TEB), and feature fusion block (FFB), each having customized functions for DF.

The three key blocks are briefly illustrated hereinafter, and later they will be reported in detail in the following subsections.
First, the SEB utilizes the spatial coherence of local climatic continuity by integrating meteorological information of neighboring regions when forecasting the drought condition of a target region, thereby improving the prediction accuracy.
Second, the TEB extracts temporal dependencies using a dual CNN-QNN framework, leveraging the state-of-the-art Mamba time sequence modeling while incorporating quantum temporal entanglement mechanisms to enhance temporal feature learning.
Specifically, TEB integrates the quantum deep network (QUEEN) \cite{hyperqueen} to capture the short-term climatic variations/dependencies, while the Mamba network efficiently learns long-range temporal patterns.
Unlike the traditional CNN that extract affine-computing features, our QUEEN adopts unitary-computing quantum neurons to extract fundamentally distinct features \cite{lin2024qmm,lin2024qhcd}; this strategy has been proven effective for diverse remote sensing applications.
By parallelly integrating classical and quantum architectures, TEB enhances the model’s expressiveness in learning intricate spatiotemporal features, leading to more robust DF.
Third, the extracted spatiotemporal features from different meteorological variables are then fused by the customized FFB, thereby generating the final DF results.

%%==================psuedo code==============%%
\begin{algorithm}[t]
\caption{The proposed SQUARE-Mamba algorithm}\label{alg:SQUARE-Mamba}
\begin{algorithmic}[1]
    \State \textbf{Given} the time-series data $\bZ\in \mathbb{R}^{\text{15} \times 7}$ over 15 months with 7 critical meteorological statistics.
    \State Vectorize $\bZ$ to $\bz\in \mathbb{R}^{105}$ for ease of the subsequent spatial augmentation.  
    \State Obtain the spatially augmented tensor $\mathcal{T}(\bz)\in \mathbb{R}^{105\times 3\times 3}$ from the neighboring $3\times 3$ regions.
    \State Compute the spatially encoded feature $\bs \in \mathbb{R}^{105}$ using the spatial encoding block $\text{SEB}(\cdot)$ based on \eqref{eq:seb}.
    \State Reshape $\bs$ back to $\bS\in\mathbb{R}^{\text{15} \times 7} $.
    \State Compute the temporally encoded features $\bF \in \mathbb{R}^{15 \times 7}$ using the temporal encoding block $\text{TEB}(\cdot)$ based on \eqref{eq:teb}.
    \State Obtain the drought condition $d \in \mathbb{R}$ using the feature fusion block $\text{FFB}(\cdot)$ based on \eqref{eq:ffb}.
    \State Standardize the $d$ value to $[-3,3]$ using the scaled Tanh function $\text{s-Tanh}(\cdot)$.
    \State \textbf{Output} the predicted drought condition $d\in[-3,3]$.
\end{algorithmic}
\end{algorithm}

%%===================SEB====================%%
\begin{figure*}[t]
\centering
\includegraphics[width=1\textwidth]{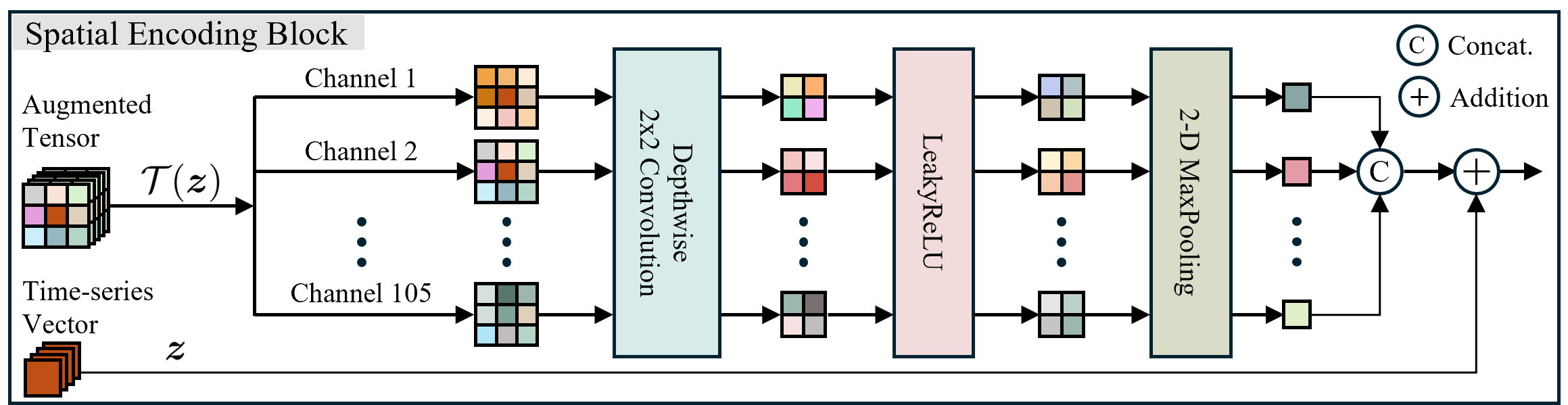}
\caption{Detailed architecture of the proposed spatial encoding block (SEB), which is designed to extract spatial features by employing the climatic continuity in the 8 adjacent areas as specified by $\mathcal{T}(\bz)$.
The 105 channels correspond to 7 different meteorological variables (i.e., precipitation, maximum/minimum/mean temperatures,
vapor pressure, cloud cover, and potential evapotranspiration) over continuous 15 months.}
\label{fig:seb}
\end{figure*}

Mathematically, according to the above illustration, the overall architecture of the SQUARE-Mamba network is defined as follows:
\begin{align}
d=\text{s-Tanh}(\text{FFB}(\text{TEB}(\text{SEB}(\bz,\mathcal{T}(\bz))))),\label{eq:framework}
\end{align}
where $d\in$ $[-3,3]$ represents the forecasting result indicating the predicted drought conditions; 
$\text{s-Tanh}(\cdot)$ denotes the scaled hyperbolic tangent function \cite{Gnanasambandamtanhpami2023} returning a value within the interval $[-3,3]$; 
$\text{FFB}(\cdot)$ denotes the feature fusion block function; 
$\text{TEB}(\cdot)$ denotes the temporal encoding block function; $\text{SEB}(\cdot)$ denotes the spatial encoding block function;
$\bz\in \mathbb{R}^{105}$ is the input time series data (at a target region) across 15 months, each having 7 meteorological variables (i.e., precipitation, maximum/minimum/mean
temperatures, vapor pressure, cloud cover, and potential evapotranspiration); 
the tensor $\mathcal{T}(\bz)\in \mathbb{R}^{105\times3\times3}$ is the spatially augmented version of the data $\bz\in \mathbb{R}^{105}$ (at the target region) by including the data from its 8 neighboring regions (cf. Figure \ref{fig:seb}).
Following a similar setting \cite{DIKSHIT2021111979}, we assume that the drought condition $d$ of a given month can be predicted from the historical meteorological data over the past $15$ months.

Based on the elegant and concise quantum/classical deep model \eqref{eq:framework}, we then construct the effective DF algorithm, i.e., Algorithm \ref{alg:SQUARE-Mamba}, which summarizes the above discussion and design philosophy for better understanding.
%by integrating three main components, which will be detailed later.
%
Our experiments will show that the spatial augmentation strategy $\mathcal{T}(\bz)$ and the QUEEN-based temporal analysis both play essential roles in yielding accurate forecasting results.
The proposed SEB, TEB, and FFB deployed in SQUARE-Mamba (i.e., Algorithm \ref{alg:SQUARE-Mamba}) will be detailed in Sections \ref{sec:DFSEB}, \ref{sec:DFTEB}, and \ref{sec:DFFFB}, respectively.

\subsection{Spatial Encoding Block} \label{sec:DFSEB}
Owing to the strong spatial correlations in meteorological data, neighboring areas tend to exhibit similar climate characteristics, as can be observed from Figure \ref{fig:study locations}.
However, existing benchmark forecasting models frequently treat each location independently, thereby neglecting crucial spatial dependencies and limiting their prediction performance.
To address this limitation, we introduce the spatial encoding block (SEB) to embed climatic spatial dependencies into the forecasting process within the proposed SQUARE-Mamba model.
By modeling spatial dependencies, our approach enables the model to more effectively capture regional correlations, thereby enhancing prediction reliability and accuracy.
Specifically, we transform the target input vector, which spans 15 months and includes the 7 meteorological variables defined above (i.e., $\bz\in\mathbb{R}^{105}$) into the 3-D augmented tensor $\mathcal{T}(\bz)\in\mathbb{R}^{105\times w\times w}$ using the operator ``$\mathcal{T}:\mathbb{R}^{105}\rightarrow\mathbb{R}^{105\times w\times w}$''.
The resulting 3-D tensor $\mathcal{T}(\bz)$ is constructed from the continuous $w\times w$ pixel region surrounding the target input vector $\bz$, where the window size $w$ is set to 3 in this work, and $\bz$ is positioned at the window center.
To address potential missing values in $\mathcal{T}(\bz)$, one can simply apply the $k$-nearest neighbor (KNN) imputation with $k:=1$ \cite{nearestpadding,lin2024superrpca} during the augmentation procedure.

As shown in Figure \ref{fig:seb}, the overall procedure of the SEB, denoted as $\text{SEB}(\cdot)$, is defined as follows:  
\begin{align}
\text{SEB}(\bz,\mathcal{T}(\bz))=\bz+\text{Maxpooling}(\sigma(f_\text{D}(\mathcal{T}(\bz)))),\label{eq:seb}
\end{align}
where $\text{Maxpooling}(\cdot)$ represents the max-pooling operator; $\sigma(\cdot)$ denotes the LeakyReLU activation function with a negative slope of 0.2 \cite{maas2013rectifier}; and $f_\text{D}(\cdot)$ is a $2 \times 2$ depthwise convolutional layer.  
The core idea of \eqref{eq:seb} is to extract spatial information from four distinct directions (i.e., corners) using depthwise convolution.  
Notably, the depthwise convolutional layer aggregates spatial dependencies while preserving essential temporal information within each channel.  
Subsequently, the most representative spatial feature is selected via the max-pooling operator, ensuring that the most salient information is retained.  
Finally, a shortcut connection mechanism is introduced to integrate the extracted adjacent spatial features with the target pixel $\bz$, thereby obtaining the spatially encoded feature $\bs\in\mathbb{R}^{105}$.  
By following \eqref{eq:seb}, SEB, which can be easily implemented, encodes spatial dependencies effectively while maintaining high computational efficiency.

\subsection{Temporal Encoding Block}\label{sec:DFTEB}
\begin{figure*}[t]
\centering
\includegraphics[width=0.99\textwidth]{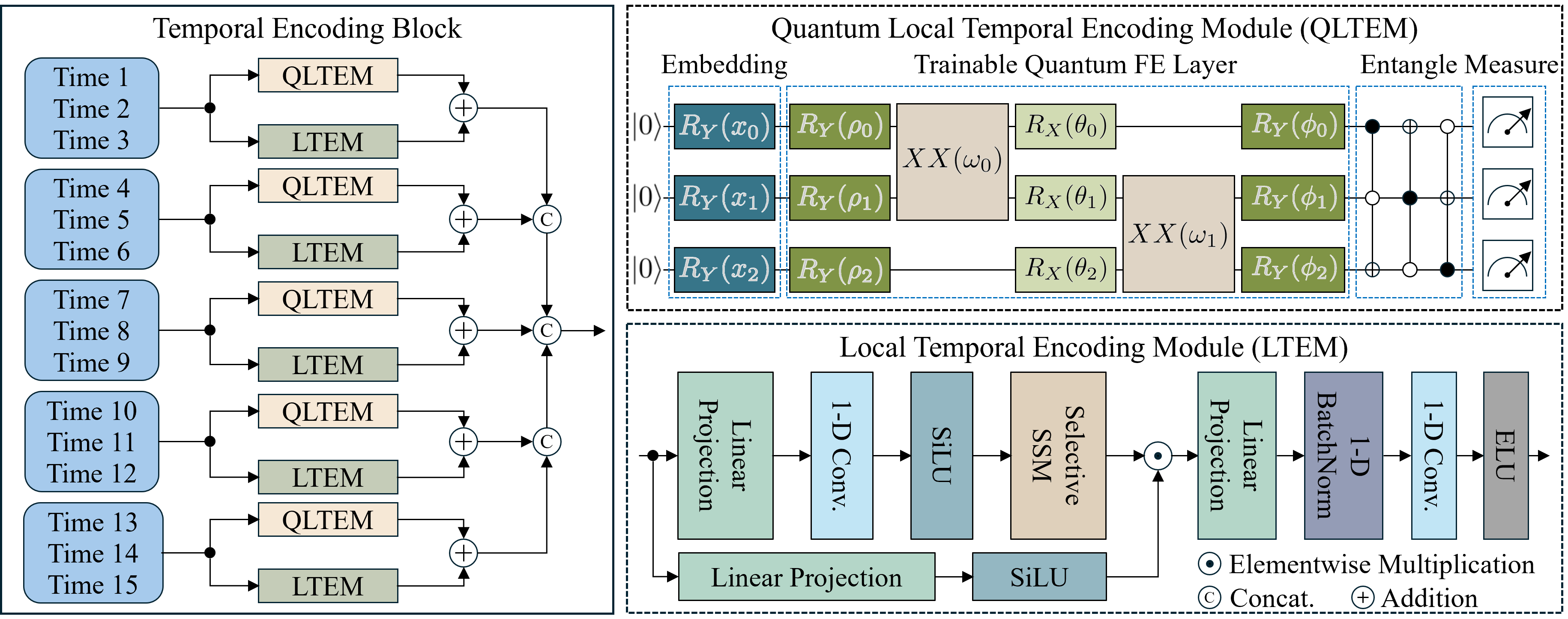}
\caption{Detailed architecture of the proposed temporal encoding block (TEB), which adopts a parallelized classical/quantum deep structure for temporal feature extraction.
Simply speaking, TEB learns temporal dependencies of the spatially encoded features, thereby yielding the exquisite ``spatiotemporal'' features.
To establish global temporal relations, we first partition the sequence into five local groups to extract short-term variations through the local time encoding module (LTEM) and Quantum LTEM (QLTEM), where the Mamba temporal modeling (used in LTEM) and the temporal quantum entanglement (used in QLTEM) greatly contribute to the learning of temporal dependencies.
These locally correlated features are then combined as complete temporal representations, allowing the model to learn both short-term fluctuations and long-term trends effectively.}
\label{fig:TEM}
\end{figure*}

As extreme climate events get more frequently occurred, short-term temperature and precipitation fluctuations have been intensified \cite{extremeclimate}.  
Therefore, accurately capturing these fluctuations is crucial for improving the predictive performance of DF.  
To address this challenge, we customize a temporal encoding block (TEB) designed to capture short-term fluctuations while effectively preserving long-term trends. 
As shown in Figure \ref{fig:TEM}, TEB models temporal dependencies across five local time groups, thereby enhancing the model’s ability to detect rapid climate variations.

Specifically, the proposed TEB function, denoted as $\text{TEB}(\cdot)$, is defined as follows:
\begin{align}
\text{TEB}_i(\cdot)&=\text{LTEM}_i(\cdot)+\text{QLTEM}_i(\cdot), ~i \in \{1,\dots,5\},\label{eq:teb}
\end{align}
where $i$ denotes the $i$th local time group in $\text{TEB}(\cdot)$; $\text{TEB}_i(\cdot)$ denotes the $i$th local temporal encoding block function; $\text{LTEM}_i(\cdot)$ represents the $i$th local time encoding module function; $\text{QLTEM}_i(\cdot)$ represents the $i$th quantum local time encoding module function.
Notably, five local groups in \eqref{eq:teb} have non-shared weights, allowing them to capture distinct features from different temporal patterns.
To preserve long-term trends, all locally encoded temporal features are concatenated into a comprehensive long-term representation for feature fusion.

Within our model design, the local time encoding module (LTEM) serves as the core component of TEB, enabling the precise modeling of short-term temporal dependencies.
By leveraging Mamba \cite{gu2023mamba}, a state-of-the-art sequence modeling method, our LTEM effectively captures temporal relationships through its strong time-series modeling capabilities.
Unlike the traditional Transformer model \cite{vaswani2017attention}, which relies on self-attention with quadratic complexity, Mamba adopts a discrete selective state-space model (SSSM) with only linear complexity, significantly improving computational efficiency, especially for long-sequence modeling.

%%====================TEM===================%%

Mathematically, the discrete SSSM function \cite{han2024demystify} (i.e., $\by=\text{SSSM}(\bx)$) can be defined as follows:
\begin{align}
\bh_i&=\overline{\bA}_i\bh_{i-1}+\overline{\bB}_ix_i,\label{eq:h}\\
y_i&=\bC_i\bh_i+Dx_i,\label{eq:y}
\end{align}
where $\bx$ and $\by\in \mathbb{R}^{L\times1}$ represent the input and output sequence vectors, respectively; $\bh_i \in \mathbb{R}^{N \times 1}$ is the $i$th hidden state; $x_i$ and $y_i \in \mathbb{R}$ denote $i$th elements of $\bx$ and $\by$, respectively;  $\overline{\bA}_i \in \mathbb{R}^{N \times N}$, $\overline{\bB}_i \in \mathbb{R}^{N \times 1}$, $\bC_i \in \mathbb{R}^{1 \times N}$, and $D \in \mathbb{R}$ are learnable weights that regulate the input $x_i$ and the hidden state $\bh_i$ in \eqref{eq:h} and \eqref{eq:y}.
If the input sequence has a length of $L$ with $K$ channels (i.e., $[\bx_1, ..., \bx_K]$), $\text{SSSM}(\cdot)$ is applied independently to each channel \cite{gu2023mamba}.
Therefore, \eqref{eq:h} and \eqref{eq:y} enable LTEM to model temporal dependencies effectively, and this is the so-called Mamba time-series modeling \cite{gu2023mamba}.
This design facilitates the climate prediction tasks involving large-scale temporal data (e.g., drought forecasting).

To be mathematically rigorous, building upon the efficiency of LTEM, we now define its architectural details as follows:
\begin{align}
 \bS_{\text{SSSM}}&=\text{SSSM}(\tau (f_1(\text{LP}_1(\bS))))\odot\tau(\text{LP}_2(\bS)),\label{eq:feature}\\
\text{LTEM}(\bS)&=\sigma(f_2(\text{BN}(\text{LP}_3(\bS_{\text{SSSM}})))),\label{eq:LTEM}
\end{align}
where $\bS_\text{SSSM}$ is the gated feature modulated by $\tau(\text{LP}_2(\bS))$; $\tau(\cdot)$ denotes the sigmoid linear unit (SiLU) activation function \cite{silu}; $f_i(\cdot)$ denotes the $i$th $3\times1$ 1-D convolutional layer along the temporal dimension; $\text{LP}_i(\cdot)$ represents the $i$th linear projection operator; $\bS\in\mathbb{R}^{15 \times 7}$ is the reshaped matrix of the spatially encoded feature vector $\bs\in\mathbb{R}^{105}$ returned by $\text{SEB}(\cdot)$; $\sigma(\cdot)$ corresponds to the exponential linear unit (ELU) activation function \cite{elu}; $\text{BN}(\cdot)$ represents the batch normalization layer.
Overall, LTEM (cf. \eqref{eq:feature} and \eqref{eq:LTEM}) leverages the SSSM to enhance temporal feature extraction, ensuring efficient and expressive sequence modeling.  

On the other hand, as alluded above, capturing/learning the short-term climatic fluctuations is critical.
So, we further incorporate the temporal entanglement among the quantum features extracted from a short time period (say, three months).
This further enhances the temporal encoding capability, as we plug the Quantum LTEM (QLTEM) into the TEB by using quantum deep network (QUEEN) \cite{hyperqueen} to learn temporal features.
Unlike classical CNN features, QUEEN is proven to be able to provide more abstract features (a.k.a., unitary features; cf. Table \ref{tab:common_qu_gate}), as having been successfully demonstrated on numerous remote sensing missions in very recent literature, including mangrove forest classification \cite{lin2024qmm}, hyperspectral change detection \cite{lin2024qhcd}, NASA optical data restoration \cite{lin2024quantum, lin2023quantum}, and underdetermined blind source separation \cite{lin2024prime}.
Specifically, QNN leverages unitary computing and quantum entanglement mechanisms \cite{hyperqueen} to enhance model's representation capacity.
This motivates us to adopt QNN to extract entangled temporal features within the dual-branch network architecture (cf. Figure \ref{fig:TEM}), aiming to upgrade the DF performance.

\begin{table}[t]
\scriptsize
\setlength{\tabcolsep}{5.5pt} 
% Default value: 6pt
\caption{Symbols and mathematical definitions for the quantum gates (all corresponding to unitary operators, according to the Schrödinger equation \cite{hyperqueen}) used in the QLTEM defined by Figure \ref{fig:TEM}.
With $\theta$ representing the learnable parameters, we use $\delta$ and $\gamma$ to denote $\cos(\frac{\theta}{2})$ and $\sin(\frac{\theta}{2})$, respectively, for conciseness.
Additionally, DIAG($A$, $B$, $C$) represents a block-diagonal matrix with diagonal blocks $A$, $B$ and $C$, and $\bI_n$ denotes the $n\times n$ identity matrix.}\label{tab:common_qu_gate}
%\vspace{-0.2cm}
\begin{center}
\renewcommand\arraystretch{1}
\begin{tabular}{|c c c|} 
 \hline
 \rule{0pt}{2ex}
 Quantum Gate & Symbol & Unitary Operator 
 \rule{0pt}{2ex}
 \\
 \hline
 \rule{0pt}{4ex}
 Rotation X
 &
 \begin{tikzcd}
    \qw & \gate{R_{X}(\theta)} & \qw
 \end{tikzcd}
 &
 $\begin{pmatrix}
    \delta & -i \gamma \\
    -i \gamma & \delta
\end{pmatrix}$
 \rule{0pt}{4ex}
 \\ 
 \hline
 \rule{0pt}{4ex}
 Rotation Y
 &
 \begin{tikzcd}
    \qw & \gate{R_{Y}(\theta)} & \qw 
 \end{tikzcd}
 & 
 $\begin{pmatrix}
    \delta & - \gamma \\
    \gamma & \delta
\end{pmatrix}$
 \rule{0pt}{4ex}
\\
 \hline
 \rule{0pt}{4ex}
 \rule{0pt}{6.5ex}
 Ising XX
&
\begin{quantikz}
    \qw & \gate{XX(\theta)} & \qw
\end{quantikz}
&
$\begin{pmatrix}
    \delta & 0 & 0 & -i\gamma\\
    0 & \delta & -i\gamma & 0\\
    0 & -i\gamma & \delta & 0\\
    -i\gamma & 0 & 0 & \delta
\end{pmatrix}$
\\
\hline
Pauli-Z &
\begin{tikzcd}
\meter{} 
\end{tikzcd}
&
$\begin{pmatrix}
    1 & 0 \\
    0 & -1 \\
\end{pmatrix}$
%\rule[-5ex]{0pt}{4ex}
 \\
 \hline
NOT &
\begin{tikzcd}
\qw & \gate{X} & \qw
\end{tikzcd}
&
$\begin{pmatrix}
    0 & 1 \\
    1 & 0 \\
\end{pmatrix}$
%\rule[-5ex]{0pt}{4ex}
 \\
 \hline
 \rule{0pt}{11ex}
Toffoli (CCNOT)
 &
 \begin{tikzcd}
    \qw & \ctrl{1} & \qw \\
    \qw & \octrl{1} & \qw \\
    \qw & \targ{} & \qw
 \end{tikzcd}
 &
 $\textrm{DIAG}(\bm{I}_4,X,\bm{I}_2)$
 \rule[-5ex]{0pt}{4ex}
 \\
 \hline
\end{tabular}
 %\vspace{-0.3cm}
\end{center}
\end{table}

%===================Implementation Table=============%
\begin{center}
\begin{table}[t]
    \caption{Implementation details of QLTEM, including the embedded core FE module (cf. Theorem \ref{theorem: FE}).
    The quantum gates $R_Y(g,q)$, $R_X(g,q)$, $Z(g,q)$, $XX(w_1,w_2)$, and CCNOT$(cw_1, cw_2, tw)$ follow the notation defined in Table~\ref{tab:common_qu_gate}, where $g$ and $q$ denote the number of parallel groups and qubits per group; $XX$ operates on wires $w_1$ and $w_2$; and CCNOT acts on the target wire $tw$ when control wires $cw_1$ and $cw_2$ are in states $\lvert 1 \rangle$ and $\lvert 0 \rangle$.
    }
    \centering
    \label{tab:configuration}
    \renewcommand{\arraystretch}{1.2}{
\begin{tabular}{c|c|c}
        \hline
        \hline
        Layer & Configuration & Output Size \\
        \hline
        \hline
        Input & - & 3$\times$7 \\
        \hline
        Reshape & - & 7$\times$3 \\
        \hline
        \makecell{Data\\Embedding} & $R_Y(7,3)$ & 7$\times$3 \\
        \hline
        Unitary Gate 1 & $R_Y(7,3)$ & 7$\times$3 \\
        \hline
        Unitary Gate 2 & $XX(0,1)$ & 7$\times$3 \\
        \hline
        Unitary Gate 3 & $R_X(7,3)$ & 7$\times$3 \\
        \hline
        Unitary Gate 4 & $XX(1,2)$ & 7$\times$3 \\
        \hline
        Unitary Gate 5 & $R_Y(7,3)$ & 7$\times$3 \\
        \hline
        \multirow{3}{*}{\makecell{Toffoli\\Entanglement}} 
            & CCNOT(0,1,2) & 7$\times$3 \\
        \cline{2-3}
            & CCNOT(1,2,0) & 7$\times$3 \\
        \cline{2-3}
            & CCNOT(2,0,1) & 7$\times$3 \\
        \hline
        \makecell{Quantum\\Measurement} & $Z(7,3)$ & 7$\times$3 \\
        \hline
        Reshape & - & 3$\times$7 \\
        \hline
        Output & - & 3$\times$7 \\
        \hline
        \hline
    \end{tabular}}
\end{table}
\end{center}     
%%%%%%%%%%%%%%%%%%%%%%%%%%%%%%%%%%%%
Simply speaking, QUEEN is to replace the neurons in the traditional deep network (e.g., CNN) by so-called quantum neurons.
Before presenting the design principles and implementation details, we first introduce the notations and mathematical definitions of the quantum neurons used in QLTEM, as summarized in Table \ref{tab:common_qu_gate}.  
Note that quantum neurons are implemented through some Hamiltonian over a specific time interval, ensuring that their corresponding operators remain unitary, as dictated by the Schrödinger equation \cite{Qunitary}.
By leveraging these quantum operations, QNNs have demonstrated their effectiveness across various applications \cite{hyperqueen,lin2024qmm, lin2024qhcd, lin2024prime}, particularly in hybrid architectures that combine classical networks and QNNs.
This convergence highlights the advantage of integrating quantum computing to tackle complicated problems.
Inspired by these promising results, we adopt a parallel structure integrating the classical NN and QNN.

The parallel structure is illustrated in Figure \ref{fig:TEM}, where we incorporate Rotation X, Rotation Y, Ising XX, Pauli-Z, and Toffoli quantum gates to develop interpretable QLTEM.
Within this framework, QLTEM extracts temporal features from each $1 \times 3$ time-wise sequence using the $R_Y$-Ising-$R_X$-Ising-$R_Y$-Toffoli$^3$ architecture, which has been empirically validated for its effectiveness (cf. Figure \ref{fig:TEM}).
Specifically, the QNN learns the quantum unitary transformation with a set of trainable parameters $\{\rho_k, \omega_k, \theta_k, \phi_k\}$.
Furthermore, by leveraging the entanglement mechanism and measurement, QLTEM captures finer temporal dependencies, thereby improving predictive performance.
In summary, the implementation details of the proposed QLTEM are presented in Table \ref{tab:configuration}.
Specifically, Figure \ref{fig:TEM} and Equation \eqref{eq:teb} illustrate how the quantum model is integrated into the SQUARE-Mamba architecture.
In Figure \ref{fig:TEM}, quantum neurons from Table \ref{tab:common_qu_gate} are embedded within the QLTEM (cf. the upper right part), which operates alongside the classical LTEM in a dual-track structure (cf. the left part).
This dual structure is motivated by the information theory \cite{lin2024qmm, lin2024qhcd}, which suggests that combining highly diverse information sources from both the classical and quantum features can facilitate the decision-making procedure.
Additionally, QLTEM is designed to achieve quantum full expressibility (FE), guaranteeing that the network can express all valid quantum functions.
This property is formally established in Theorem \ref{theorem: FE} below.
\begin{theorem} \label{theorem: FE}
The trainable quantum neurons deployed in the proposed QLTEM (cf. Figure \ref{fig:TEM}) can express any valid quantum unitary operator $U$, parameterized by a set of real-valued trainable network variables $\{\rho_k, \omega_k, \theta_k, \phi_k\}$.
\hfill$\square$
\end{theorem}

\noindent
The proof of Theorem \ref{theorem: FE} can be done by utilizing the transparency of the Ising quantum neurons.
The exact proof procedure follows a similar approach to that of \cite[Theorem 2]{hyperqueen} and is omitted here for conciseness.
To conclude, the TEB (cf. \eqref{eq:teb}) can be effectively implemented based on the aforementioned structure.

\subsection{Feature Fusion Block}\label{sec:DFFFB}
%
%%===========feature fusion block===========%%
\begin{figure}[t]
\centering
\includegraphics[width=0.46\textwidth]{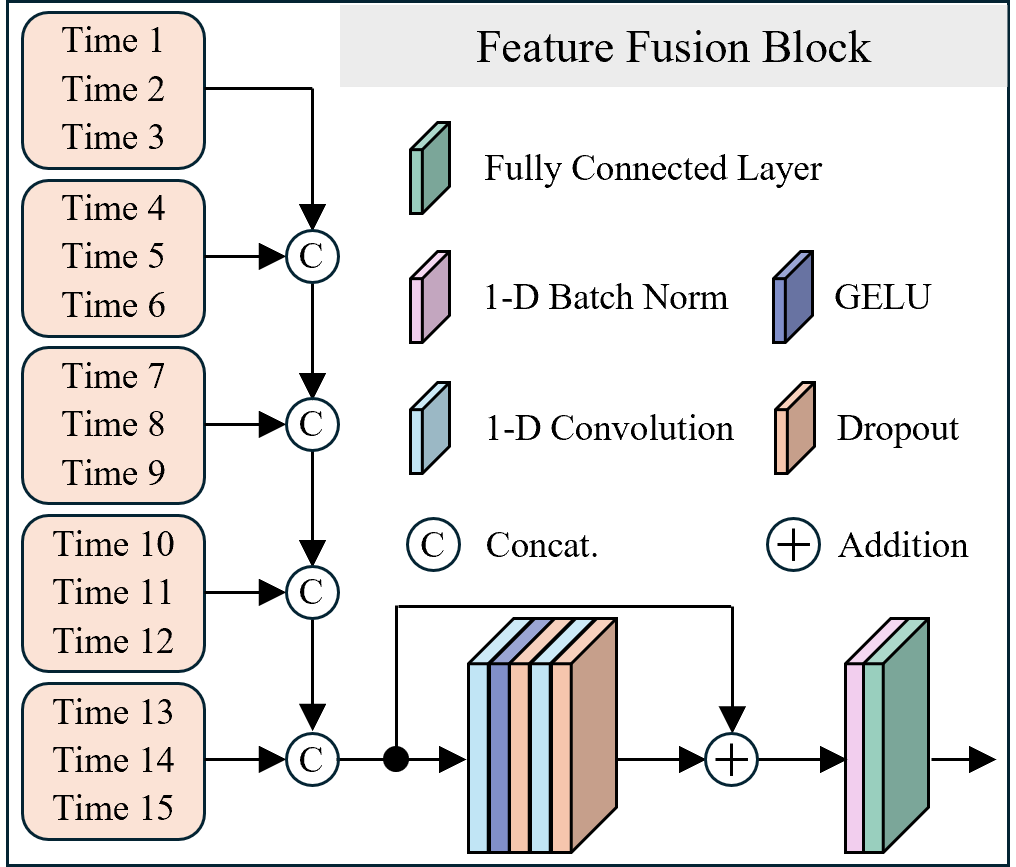}
\caption{Detailed architecture of the proposed feature fusion block (FFB).
The proposed FFB aims to fuse the spatiotemporally encoded features of the seven meteorological variables (i.e., precipitation, maximum/minimum/mean temperatures,
vapor pressure, cloud cover, and potential evapotranspiration) to predict the final drought condition of the next month (i.e., the month next to the input 15 continuous months).}
\label{fig:ffb}
\end{figure}

Once the LTEM and QLTEM jointly capture temporal features of the spatial characteristics returned by SEB, the proposed SQUARE-Mamba has extracted the exquisite spatiotemporal features from different meteorological variables.
These spatiotemporal representations are then fused to maximize the predictive accuracy for effective DF.
To fully leverage the complementary strengths of these two modules, we introduce the feature fusion block (FFB), which integrates the temporal information from classical LTEM and quantum QLTEM encoding processes to generate the final prediction result.
Specifically, FFB aggregates all encoded spatiotemporal features, as illustrated in Figure \ref{fig:ffb}.

The mathematical formulation of the module $\text{FFB}(\cdot)$ is defined as follows:
\begin{align} 
\text{FFB}(\bF)&=\text{FCL}(\text{BN}(\bF+\bF')),\label{eq:ffb}\\
\bF'&=\text{Drop}_2(f_2(\text{Drop}_1(\sigma(f_1(\bF))))),\label{eq:fdrop} 
\end{align} 
where $\text{FCL}(\cdot)$ denotes a fully connected layer;
$\bF\in\mathbb{R}^{15 \times 7}$ is the temporally encoded features obtained through $\text{TEB}(\cdot)$;
$\bF'\in\mathbb{R}^{15 \times 7}$ represents the embedded features;
$\text{Drop}_i(\cdot)$ refers to the $i$th dropout layer with a dropping probability of 0.2 \cite{dropout};
$f_i(\cdot)$ is the $i$th $1\times1$ 1-D convolutional layer applied along the temporal dimension;
and $\sigma(\cdot)$ denotes the Gaussian error linear unit (GELU) activation function \cite{gelu}.

The core function of FFB (cf. \eqref{eq:ffb} and \eqref{eq:fdrop}) is to refine and effectively fuse spatiotemporal feature representations, ensuring the integration of complementary classical-QNN information.
By incorporating the dropout mechanism, FFB selectively filters and propagates informative features to deeper layers, mitigating overfitting and improving generalization.
In addition, a shortcut connection mechanism is employed to preserve the original spatiaotemporally encoded features, ensuring that critical information is retained throughout the fusion process.
Finally, the fully connected layer (FCL) produces the final prediction, demonstrating the effectiveness of the fusion mechanism in integrating valuable information.
By following \eqref{eq:ffb} and \eqref{eq:fdrop}, FFB can be efficiently implemented.

To conclude this section, we present SQUARE-Mamba, an SPEI-driven quantum spatially-aware Mamba network, as summarized in Figure \ref{fig:DFframework}.
This framework effectively leverages neighboring climatic coherence information through the SEB to enhance DF performance.
Furthermore, the integration of the computationally efficient Mamba network and the explainable QNN with FE (cf. Theorem \ref{theorem: FE}) successfully leads to rich temporal feature extraction results, well capturing both classical and quantum characteristics.
Additionally, the parallel structure of TEB ensures that both Mamba- and QNN-based features are fairly fused in the final prediction outcome.
Finally, the spatiotemporally encoded features for different meteorological variables are fused by FFB, in order to enhance predictive accuracy and also to ensure effective integration of complementary classical and quantum information.
We will conduct systematic ablation studies to evaluate the effectiveness of the spatial augmentation strategy and the temporal quantum entanglement mechanism.
We will also demonstrate the superior forecasting performance of the overall SQUARE-Mamba architecture (i.e., Algorithm \ref{alg:SQUARE-Mamba}), validating our design philosophy and highlighting its practical applicability.

\section{Experimental Results}\label{sec:experiment}
\begin{figure*}[t]
\centering
\includegraphics[width=0.9\textwidth]{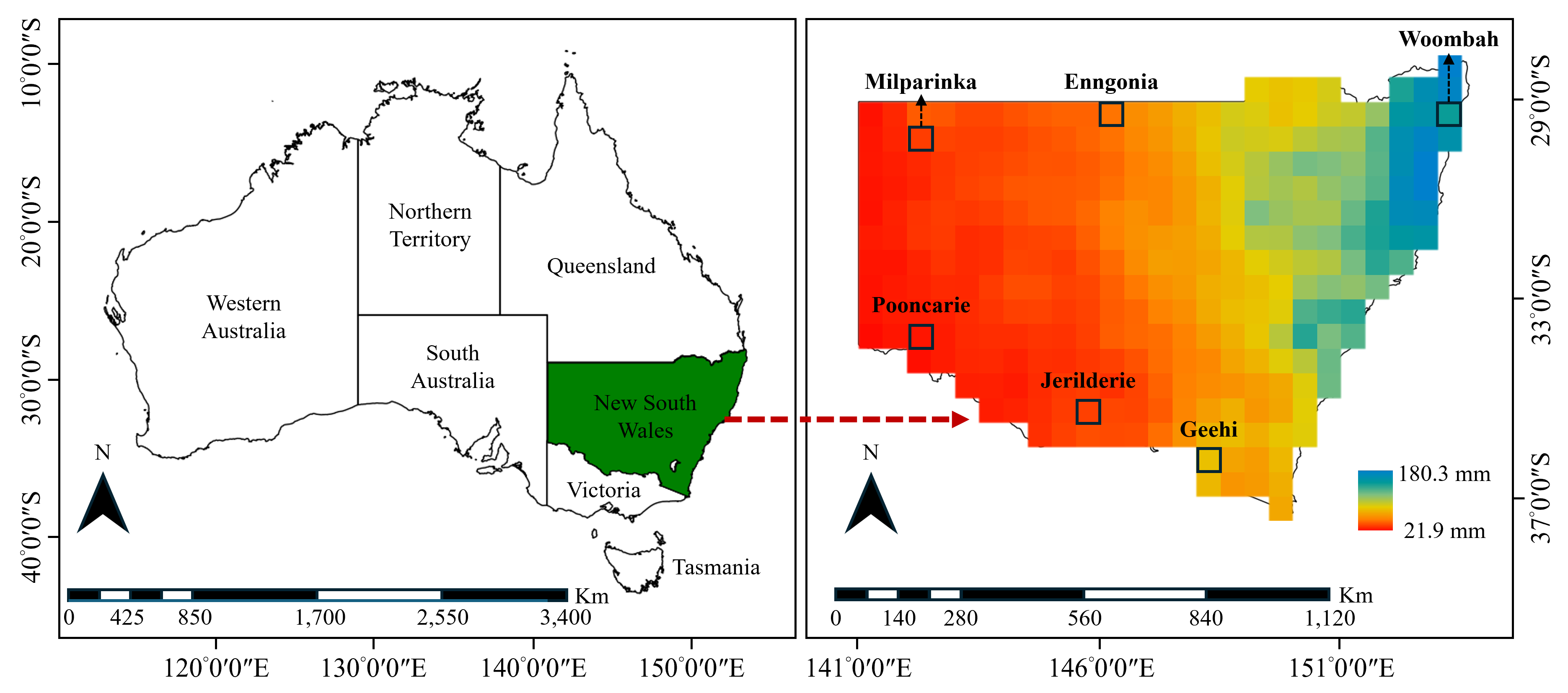}
\caption{{Investigated area in the New South Wales (NSW), Australia, and the long-term mean precipitation map using the 1961–1990 base period. 
}}
\label{fig:study locations}
\end{figure*}

In this section, we first introduce the experimental settings and then present the DF performance of SQUARE-Mamba across six representative locations in New South Wales (NSW), Australia, covering wet, dry, and moderate climatic conditions.
The testing locations include Woombah, Geehi, Enngonia, Jerilderie, Milparinka, and Pooncarie, as illustrated in Figure \ref{fig:study locations}.
Section \ref{sec:experimentsetting} provides details on the dataset description, the calculation of SPEI, the preprocessing steps, as well as the hyperparameter configurations.
Section \ref{subsec:exp} presents the forecasting results for six representative locations over a comprehensive testing period from December 2007 to December 2023, covering more than 16 years of diverse climatic conditions.
This long-term evaluation encompasses various climatic phases, including El Niño, La Niña, and normal years, ensuring a thorough assessment of model robustness under varying drought patterns.
All results are evaluated using both objective quantitative metrics and visual qualitative assessments.
In Section \ref{subsec:DFablation_study}, ablation studies are conducted to systematically analyze the effectiveness of the spatial-aware mechanism and the temporal quantum entanglement mechanism in facilitating the DF performance.
In Section \ref{sec:Performance_Insights}, we comprehensively evaluate SQUARE-Mamba, assess the computational complexity, conduct the Shapley additive explanations (SHAP) to identify the key meteorological variables, and discuss future research lines, respectively.

%=======================================%

\subsection{Experimental Settings}\label{sec:experimentsetting}

\textit{1) Data Description}: This study utilizes the Climatic Research Unit gridded time series dataset (CRU TS Version 4.08), which is produced by the University of East Anglia and is publicly available\footnote{\scriptsize\url{https://spei.csic.es/database.html}}.
Specifically, this dataset has a spatial resolution of $0.5^\circ$ latitude by $0.5^\circ$ longitude, with monthly temporal resolution over the period from 1901 to 2023 \cite{harris2020version}.
Following recent literature\cite{DIKSHIT2021111979}, seven climatic variables are selected to predict drought severity. 
These include primary variables (mean temperature and precipitation), secondary variables (vapor pressure and cloud cover), and derived variables (potential evapotranspiration, maximum temperature, and minimum temperature).
In this work, we utilize 15 months of data along with 7 meteorological variables (i.e., $\bz\in\mathbb{R}^{15 \times 7}$) to predict drought condition $d\in[-3,3]$, as summarized in Algorithm \ref{alg:SQUARE-Mamba}.
Moreover, the widely used CRU dataset has been applied to various research fields, including studies on climate variability, paleoclimatology \cite{nagavciuc2019stable}, and agronomy \cite{renard2019national}. 
%%
%%====================SPEI Description======================%%

\textit{2) Standardized Precipitation Evapotranspiration Index}: 
In SQUARE-Mamba, we use the SPEI \cite{vicente2010multiscalar} to predict drought conditions \cite{DIKSHIT2021111979, shang2023application}.
To enhance understanding, Table \ref{tab:spei table} presents the drought classifications corresponding to different SPEI values.
SPEI is derived from the climatic water balance, which is calculated as the difference between precipitation and PET.
As a result, SPEI serves as an effective metric for drought assessment.
Notably, predicting drought conditions helps mitigate significant economic losses, while meteorological drought often precedes agricultural and hydrological drought.
Thus, shorter time scales are more suitable for assessing meteorological drought than medium or longer time scales \cite{DIKSHIT2021111979}.
Given these considerations, this study focuses on predicting SPEI at a one-month time scale (i.e., SPEI-1) to facilitate practical applications.
For a detailed explanation of SPEI calculations, interested readers may refer to the recent literature \cite{vicente2010multiscalar}.

%%==========================SPEI Table======================%%
\begin{table}[t]
\caption{Drought categories based on the DF result $d$ \cite{rhee2017meteorological}.}
\renewcommand\arraystretch{1.2}
%\vspace{-0.1cm}
\begin{center}
\setlength{\tabcolsep}{0.5cm}
\scalebox{0.99}{
\begin{tabular}{c c}
\hline
\hline
\makecell[c]{Category} & \makecell[c]{Drought Forecasting Result $d$} 
\\
\hline
Extremely Dry & $~~-3\leq d\leq -2$ \\
%\hline
Severely Dry & $~~~~-2 < d \leq -1.5$\\ 
%\hline
Moderately Dry & $-1.5 < d \leq -1$\\
%\hline
Near Normal & $-1 < d \leq 1$\\
%\hline
Moderately Wet & $~~~~1 < d \leq 1.5$\\
%\hline
Severely Wet & $1.5 < d \leq 2$\\
%\hline
Extremely Wet & $~~2 < d \leq 3$\\
\hline
\hline
\end{tabular}
}
\label{tab:spei table}
\end{center}
\end{table}
%%==========================================================%%

\textit{3) Data Splitting and Preprocessing}: To ensure a fair comparison and robust evaluation, we partition the CRU dataset (1901-2023) into non-overlapping training, validation, and testing datasets \cite{belayneh2014long}.
By splitting into three disjoint datasets, one can ensure that all the methods are evaluated on completely new data, thus providing a more reliable assessment of their generalizability and effectiveness \cite{codemm}.
Accordingly, this study empirically employs about 65$\%$ (1901–1980) of the data for training, 20$\%$ (1981–2005) for validation, and the remaining portion (2006–2023) for testing.
We adopt this splitting method to ensure fair and reliable model evaluation, avoiding overly optimistic results in real-world applications.
Our primary goal is to train a model that remains robust to climate variability rather than one that merely fits historical trends.
For the time series regression task, ensuring that the input data is stationary is crucial \cite{zhang2005neural}. 
By maintaining statistical stability within each segment (i.e., 15-month intervals in this study), the model can better learn meaningful features while reducing the impact of distribution shifts across climate regimes, thus enhancing the stability of the training phase \cite{wang2024deep}.

\textit{4) Computer Facility and Hyperparameter for Training}: For training the proposed spatial-temporal drought forecasting network (i.e., SQUARE-Mamba), we employ the AdamW optimizer \cite{loshchilov2018decoupled} with a learning rate decay mechanism. 
The initial learning rate is $1\times 10^{-3}$ under the cosine annealing scheme. 
The mean squared error (MSE) loss function is utilized for the training stage.  
The training process ends once the validation loss stabilizes, typically within 250 epochs.
To evaluate the performance, we choose the checkpoint model with the best R-squared measure ($\text{R}^2$) among the validation data for the following testing experiments. 
Overall, the hyperparameter settings are robust and empirically efficient.
In addition, the details of the computational setup and resources are outlined below.
All experiments are conducted on a desktop computer with an NVIDIA Geforce RTX 4090 GPU and an Intel Core i9-13900 CPU (5.6 GHz) with 32 GB of RAM, while the numerical computing environment is Python 3.10.13 and PennyLane 0.38.0.
Specifically, we employ PennyLane \cite{pennylane}, an efficient open-source framework for simulating, training, and testing quantum circuits. 
PennyLane integrates seamlessly with various quantum hardware backends and has been widely adopted \cite{pandey2024deepfake, yu2024application}.
Notably, the pre-trained gate parameters can be deployed on real quantum hardware, such as IBMQ devices \cite{alvarez2018quantum}, as detailed and discussed in  \cite{haghparast2024innovative}.

%=======================================%
\begingroup
\setlength{\tabcolsep}{6pt} % Default value: 6pt
\renewcommand{\arraystretch}{1.2} % Default value: 1
\begin{table*}[t]
\begin{center}
% \scriptsize
\caption{Quantitative comparison using six representative locations across NSW, Australia, covering different climatic conditions (i.e., wet, moderate, and dry). A boldfaced number indicates the best performance for each location and index.}
\label{tab:quantitative}
\scalebox{0.9}{
\begin{tabular}{|cc|ccccccc|}
% \hline
\hline
Location & Index & ARIMA & LSTM & ConvLSTM & ViT & Informer & S-Mamba & SQUARE-Mamba\\
\hline
\multirow{3}{*}{Woombah} & MAE($\downarrow$) & 0.8790 & 0.3290 & 0.2888 & 0.3497 & 0.2248 & 0.3006 & {\bf0.1663}\\
& RMSE($\downarrow$) & 1.0306 & 0.4932 & 0.3835 & 0.4563 & 0.3086 & 0.4053 & {\bf0.2250}\\
& $\text{R}^2$($\uparrow$) & 0.0277 & 0.7772 & 0.8653 & 0.8094 & 0.9128 & 0.8497 & {\bf0.9536}\\
\hline
\multirow{3}{*}{Geehi} & MAE($\downarrow$) & 0.7746 & 0.4242 & 0.3249 & 0.3221 & 0.3635 & 0.4113 & {\bf0.1553}\\
& RMSE($\downarrow$) & 0.9214 & 0.5562 & 0.4493 & 0.4199 & 0.5327 & 0.5445 & {\bf0.2108}\\
& $\text{R}^2$($\uparrow$) & 0.0914 & 0.6689 & 0.7839 & 0.8113 & 0.6962 & 0.6827 & {\bf0.9524}\\
\hline
\multirow{3}{*}{Enngonia} & MAE($\downarrow$) & 0.6402 & 0.3332 & 0.2034 & 0.2527 & 0.2441 & 0.2232 & {\bf0.1948}\\
& RMSE($\downarrow$) & 0.7860 & 0.4104 & 0.2957 & 0.3314 & 0.3202 & 0.3004 & {\bf0.2867}\\
& $\text{R}^2$($\uparrow$) & 0.1524 & 0.7689 & 0.8800 & 0.8493 & 0.8593 & 0.8762 & {\bf0.8872}\\
\hline
\multirow{3}{*}{Jerilderie} & MAE($\downarrow$) & 0.6695 & 0.2187 & 0.2515 & 0.3397 & 0.2414 & 0.2564 & {\bf0.2007}\\
& RMSE($\downarrow$) & 0.8101 & 0.3149 & 0.3359 & 0.4528 & 0.3385 & 0.3618 & {\bf0.2915}\\
& $\text{R}^2$($\uparrow$) & 0.1188 & 0.8668 & 0.8485 & 0.7246 & 0.8461 & 0.8242 & {\bf0.8858}\\
\hline
\multirow{3}{*}{Milparinka} & MAE($\downarrow$) & 0.6133 & 0.3092 & 0.3601 & 0.3502 & 0.3230 & 0.2861 & {\bf0.2407}\\
& RMSE($\downarrow$) & 0.8033 & 0.4090 & 0.4650 & 0.4397 & 0.4224 & 0.3806 & {\bf0.3234}\\
& $\text{R}^2$($\uparrow$) & 0.1437 & 0.7779 & 0.7129 & 0.7434 & 0.7631 & 0.8078 & {\bf0.8612}\\
\hline
\multirow{3}{*}{Pooncarie} & MAE($\downarrow$) & 0.8002 & 0.2515 & 0.3317 & 0.3479 & 0.2990 & 0.3002 & {\bf0.1891}\\
& RMSE($\downarrow$) & 0.9697 & 0.3547 & 0.4592 & 0.4622 & 0.3910 & 0.4491 & {\bf0.2677}\\
& $\text{R}^2$($\uparrow$) & 0.0624 & 0.8745 & 0.7897 & 0.7869 & 0.8475 & 0.7989 & {\bf0.9285}\\
\hline

\end{tabular}}
\end{center}
\end{table*}
\endgroup
%=======================================================%
%
\subsection{Quantitative and Qualitative Analyses}
\label{subsec:exp}
In this section, we evaluate the effectiveness of our SQUARE-Mamba by comparing it with benchmark methods, including the autoregressive integrated moving average model (ARIMA) \cite{box2015time}, the long short-term memory network (LSTM) \cite{hochreiter1997long, DIKSHIT2021111979}, the convolutional LSTM (ConvLSTM) network \cite{ConvLSTM}, vision transformer (ViT) \cite{dosovitskiy2021an}, the transformer-based time series prediction model (Informer) \cite{zhou2021informer, shang2023application}, and Mamba-based method (S-Mamba) \cite{simplemamba}.
To ensure the robustness of these methods, we conduct experiments over long-term periods across a wide range of climatic conditions.
This comprehensive evaluation encompasses normal, El Niño, and La Niña years, allowing us to assess the reliability of studied methods under varying environmental scenarios.
Specifically, we conducted experiments over the period from December 2007 to December 2023, including normal years (i.e., 2012--2014 and 2019--2020), El Ni\~{n}o periods (i.e., 2009--2010, 2014--2016, and 2018--2019), and La Ni\~{n}a phases (i.e., 2007--2009, 2010--2012, 2016--2018, and 2020--2023).

\begin{figure*}[t]
\centering
\includegraphics[width=1\textwidth]{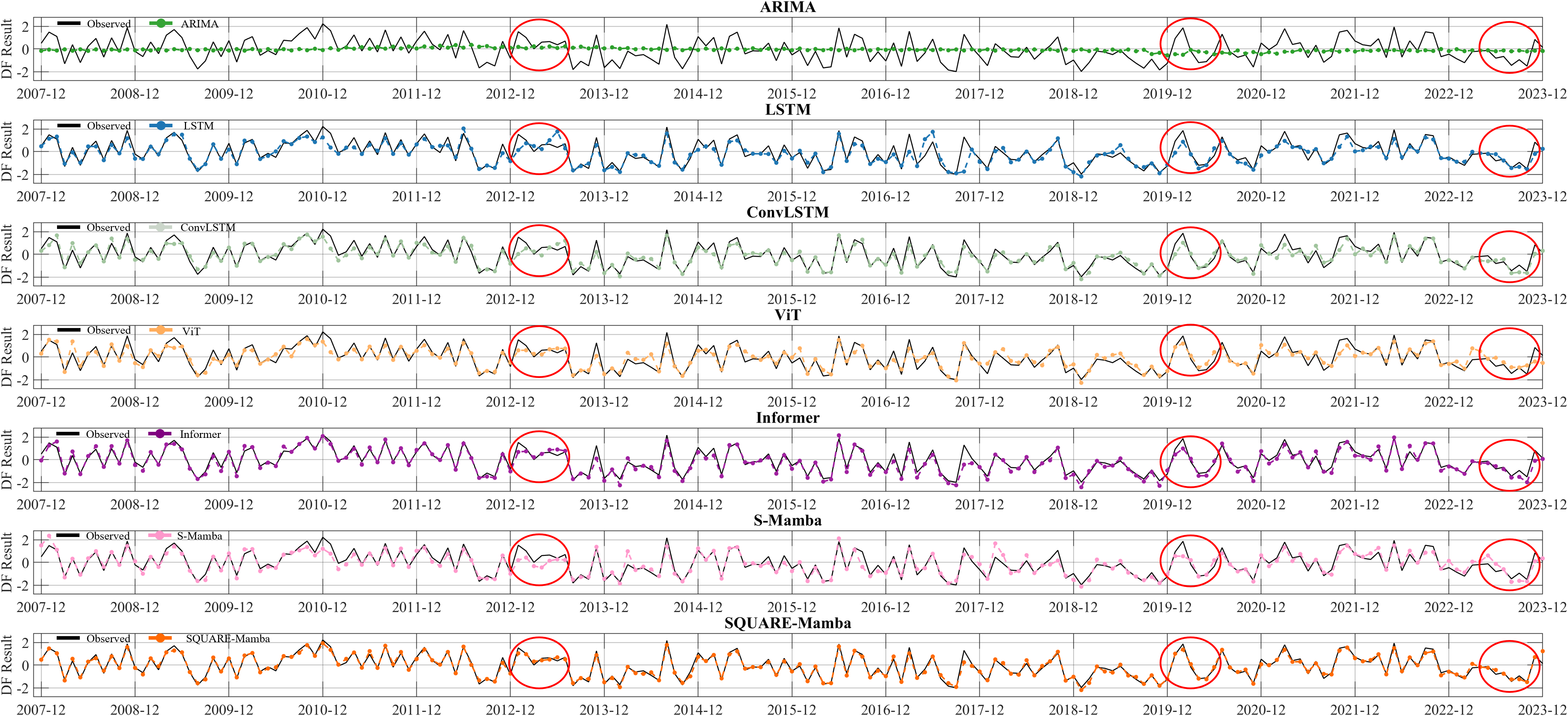}
\caption{Prediction curves (2007-2023) using various DF methods with Woombah data acquired over wet climate region.}
\label{fig:wet_1}
\end{figure*}

\begin{figure*}[t]
\centering
\includegraphics[width=1\textwidth]{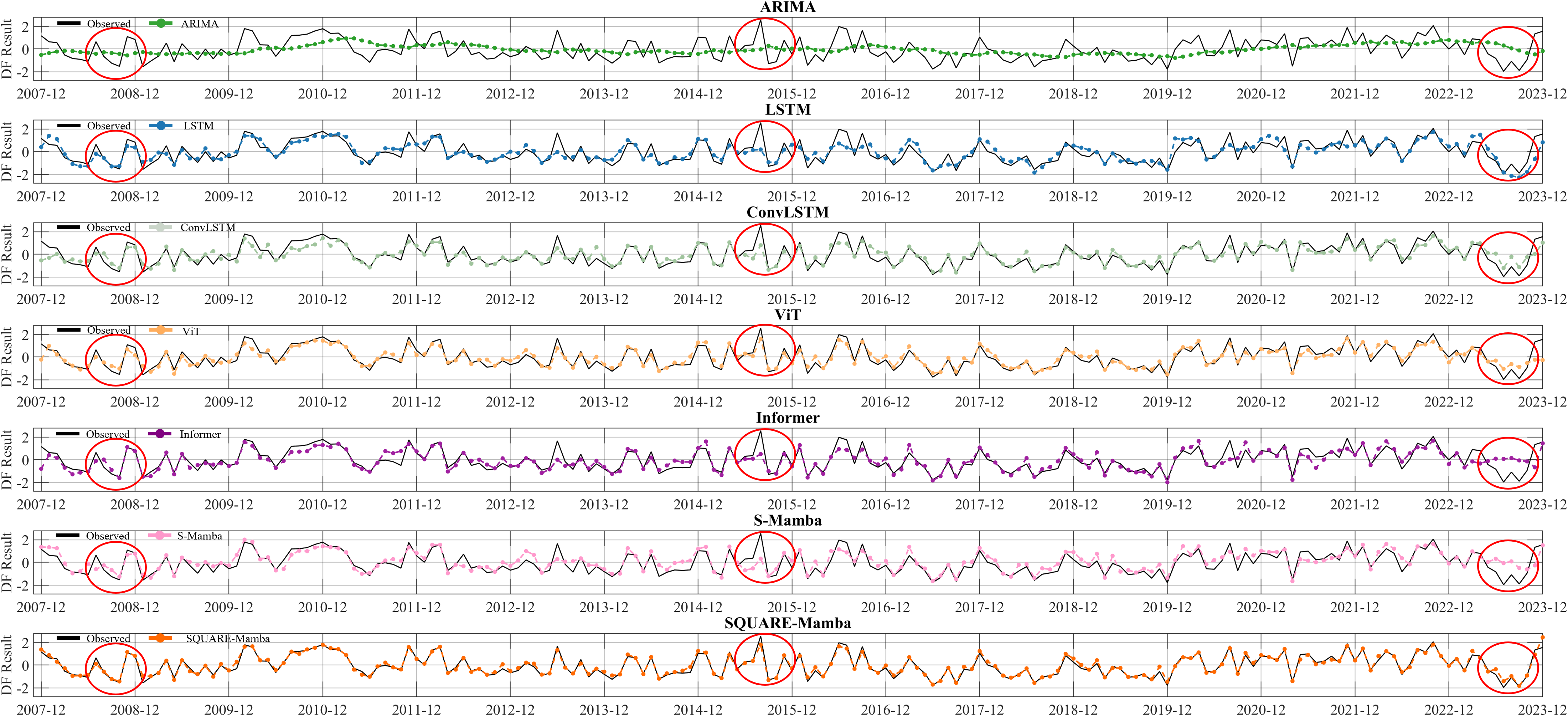}
\caption{Prediction curves (2007-2023) using various DF methods with Geehi data acquired over wet climate region.}
\label{fig:wet_2}
\end{figure*}

\begin{figure*}[t]
\centering
\includegraphics[width=1\textwidth]{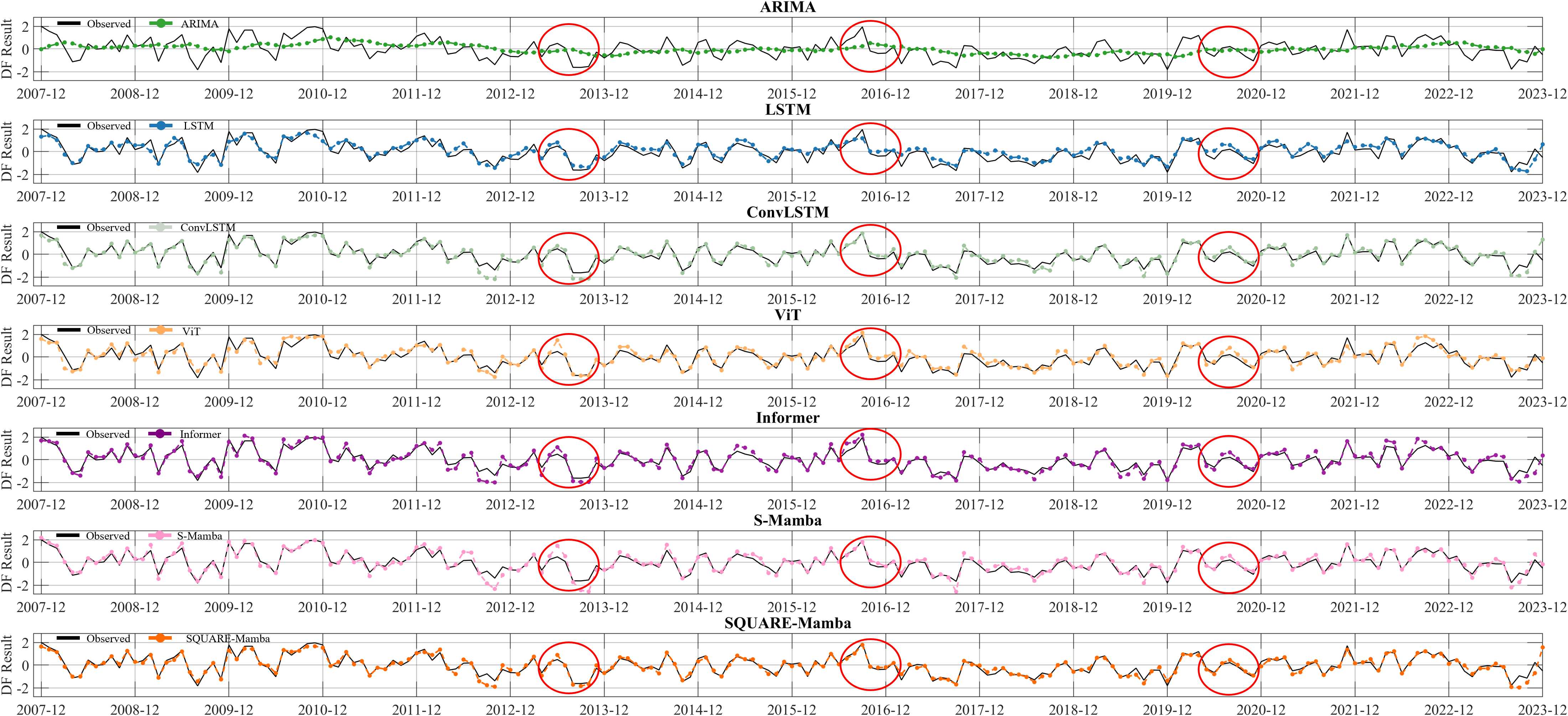}
\caption{Prediction curves (2007-2023) using various DF methods with Enngonia data acquired over moderate climate region.}
\label{fig:moderate_1}
\end{figure*}

\begin{figure*}[t]
\centering
\includegraphics[width=1\textwidth]{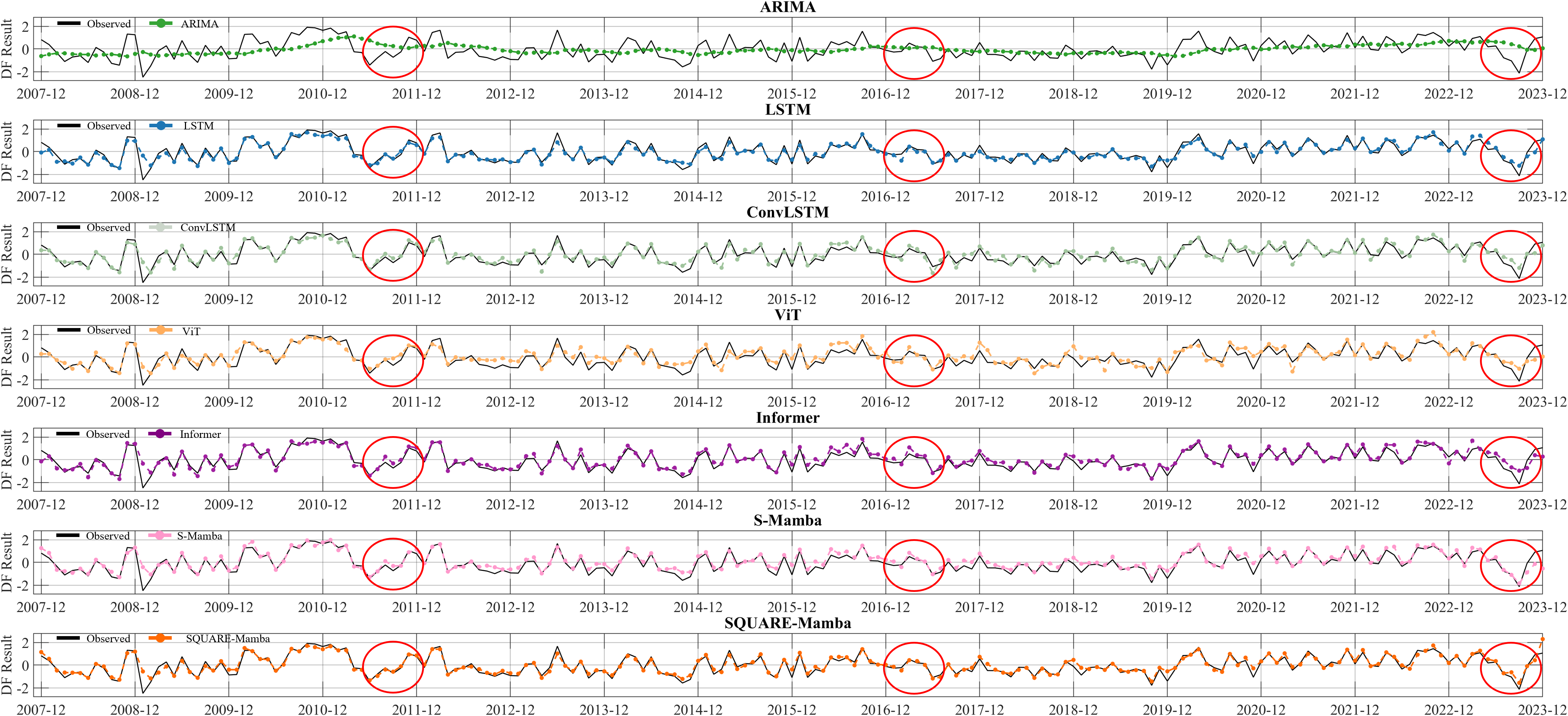}
\caption{Prediction curves (2007-2023) using various DF methods with Jerilderie data acquired over moderate climate region.}
\label{fig:moderate_2}
\end{figure*}

\begin{figure*}[t]
\centering
\includegraphics[width=1\textwidth]{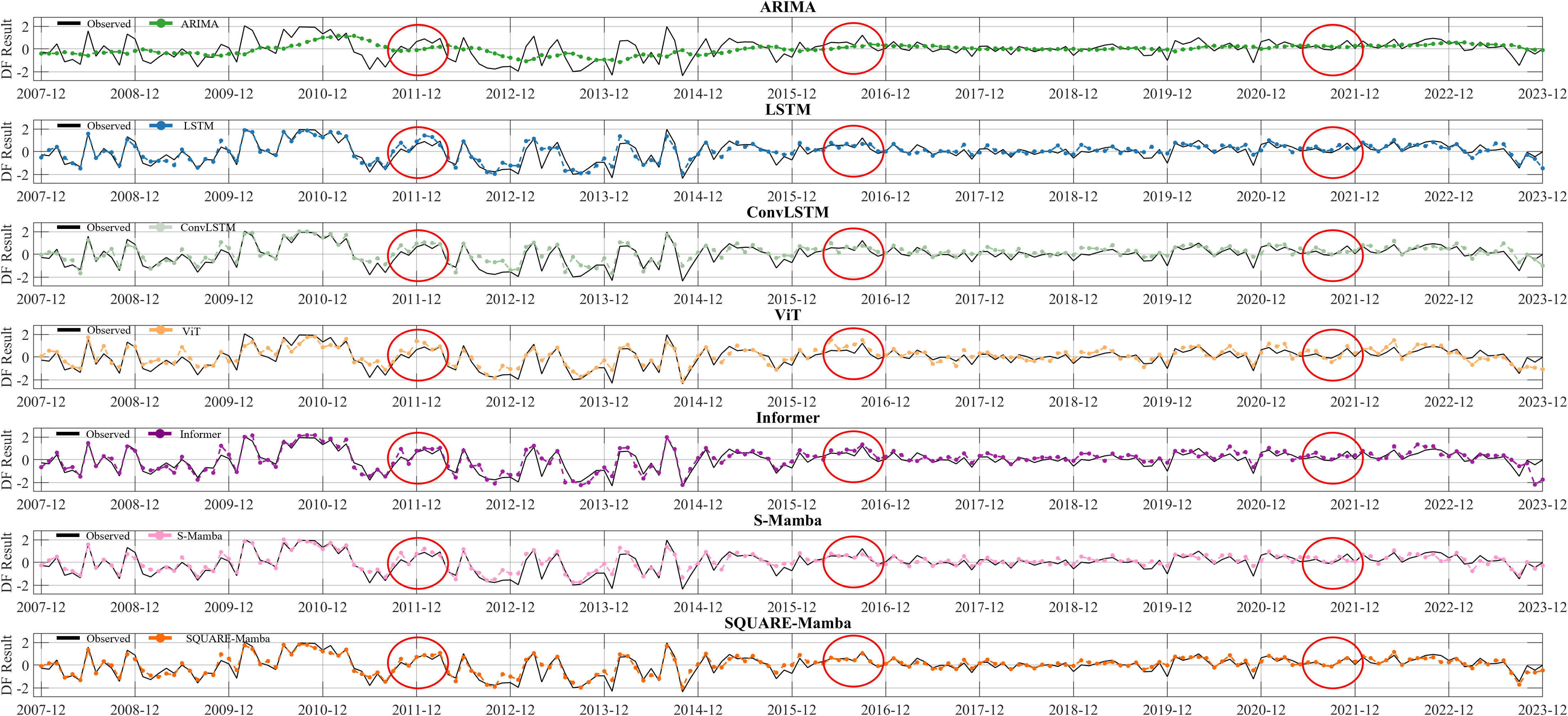}
\caption{Prediction curves (2007-2023) using various DF methods with Milparinka data acquired over dry climate region.}
\label{fig:dry_1}
\end{figure*}

\begin{figure*}[t]
\centering
\includegraphics[width=1\textwidth]{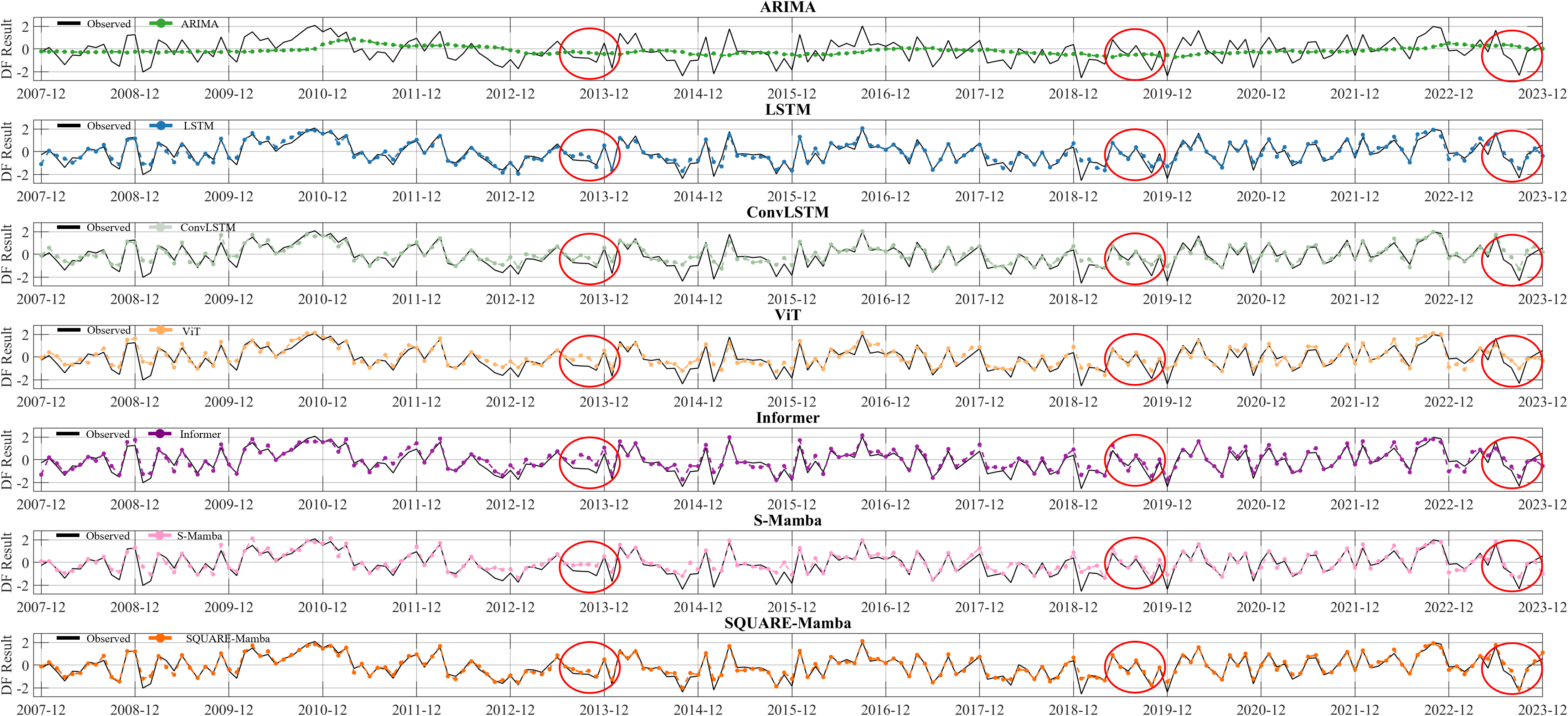}
\caption{Prediction curves (2007-2023) using various DF methods with Pooncarie data acquired over dry climate region.}
\label{fig:dry_2}
\end{figure*}

Besides evaluating performance over long-term periods, we also assess the studied methods across six representative locations that exhibit distinct climatic conditions, including wet (e.g., Woombah and Geehi), moderate (e.g., Enngonia and Jerilderie), and dry (e.g., Milparinka and Pooncarie) regions.
As demonstrated in Figure \ref{fig:study locations}, these investigated regions are located in NSW, Australia, which has historically experienced multiple severe droughts, such as the Federation Drought (1895--1902) and the Millennium Drought (2001--2010) \cite{dikshit2020short}. 
This makes NSW an ideal testing region, as it provides diverse climatic scenarios and meets the practical need for reliable DF.
Finally, the experimental results, categorized under wet, dry, and moderate climatic conditions over the extended study period, are presented in Table \ref{tab:quantitative} and Figures \ref{fig:wet_1} to \ref{fig:dry_2}, which will be discussed later.

For quantitative assessment, three commonly used objective metrics are adopted to assess the performance of the studied methods, including mean absolute error (MAE) \cite{willmott2005advantages}, root MSE (RMSE) \cite{chai2014root}, and $\text{R}^2$ \cite{cameron1997r}.
%===================
The MAE index assesses the mean of the absolute error between predicted and observed values.
%===================
The RMSE index calculates the root mean squared error between the observed and predicted values.
Since the squaring operation amplifies the values of greater deviations, RMSE places more emphasis on larger errors.
Hence, the RMSE index is more responsive to extreme error cases.
%===================
The $\text{R}^2$ computes the coefficient of determination, defined as follows:
\begin{equation*}
\begin{aligned}
\text{R}^2 &= 1-\frac{\sum_{n=1}^N (\widehat{y}_n-y_n)^2}{\sum_{n=1}^N (\overline{y}_n-y_n)^2}, \\
\end{aligned}
\end{equation*}
\\
where $\overline{y}_n = \frac{1}{N}\sum_{n=1}^Ny_n$.
The sum of squares total (SST) indicates the sum of squared differences between individual data $y_n$ and the mean $\overline{y}_n$.
In summary, $\text{R}^2$ measures how well the model fits the data. 
Hence, the model can better explain the trends in the data if it has a higher $\text{R}^2$ score.
As the value approaches $1$, the model explains most of the data's variability.
%======================================%
%=======================================%

The three objective metrics are utilized to assess the DF performance across six representative locations, as presented in Table \ref{tab:quantitative}.
For clarity, the best results are highlighted in boldfaced numbers, where lower MAE/RMSE values and higher R$^2$ scores indicate better DF performances.
As expected, the proposed SQUARE-Mamba achieves state-of-the-art results, owing to its spatially-aware mechanism that effectively captures spatial coherence (i.e., SEB), its Mamba-based time sequence modeling, as well as its integration of the quantum deep network (QUEEN).
In contrast, other benchmark models exhibit noticeable misalignment between predicted and observed trends, resulting in higher DF errors across the evaluation metrics, as will be discussed below.

%=======================================%
%=======================================%
%=======================================%
First, we conduct experiments in wet climatic regions (i.e., Woombah and Geehi), as shown in Figures \ref{fig:wet_1} and \ref{fig:wet_2}, where the first to fourth rows represent the prediction curves of ARIMA, LSTM, Informer, and SQUARE-Mamba, respectively.
Compared to the prediction curves of SQUARE-Mamba, other benchmark methods exhibit noticeable deviations from the observed curves (i.e., ground-truth curves), particularly at the time points highlighted by red-marked circles.
As shown in Figure \ref{fig:wet_1}, the rightmost circle highlights the superior forecasting capability of SQUARE-Mamba even during the El Niño event in 2023.
Most importantly, the year 2023 experienced an abrupt and intense climate shift within a short period. 
During this period, La Niña conditions persisted in January, followed by a normal phase in February, March, and April, before rapidly transitioning into an El Niño event in May 2023, which lasted until April 2024.
Despite these rapid climatic changes, SQUARE-Mamba still obtains satisfactory prediction results, thanks to the quantum/classical-hybrid temporal modeling of TEB (cf. Figure \ref{fig:TEM}), which effectively captures short-term fluctuations.
Furthermore, the middle circle in Figure \ref{fig:wet_2} demonstrates that SQUARE-Mamba delivers more accurate DF results among studied methods during the 2014--2016 El Niño event.
Notably, the 2014–2016 El Niño event has been documented as one of the most significant since records\footnote{\scriptsize\url{https://origin.cpc.ncep.noaa.gov/products/analysis_monitoring/ensostuff/ONI_v5.php}} began in 1950, with the Oceanic Niño Index (ONI) \cite{oni} exceeding 2$^\circ$C at its peak during this period.
As shown in the first and second rows of Table \ref{tab:quantitative}, the quantitative metrics double confirm the superiority of SQUARE-Mamba over benchmark methods, with $\text{R}^2$ remarkably exceeding 0.9 in both Woombah and Geehi.
%=======================================%
%=======================================%
%=======================================%

Second, we analyze the moderate climatic regions (i.e., Enngonia and Jerilderie), which pose greater challenges for DF evaluation compared to the wet areas due to their lower precipitation, as shown in Figures \ref{fig:moderate_1} and \ref{fig:moderate_2}.
As expected, SQUARE-Mamba consistently outperforms benchmark methods in DF.
As shown in the middle circle of Figures \ref{fig:moderate_1}, the SQUARE-Mamba demonstrates satisfactory DF results even during the La Niña event in 2016.
Similar to the short-term fluctuation in 2023, the year 2016 also experienced a sudden and intense climate transition within a short timeframe. 
During this period, El Niño persisted through April, shifted to normal conditions from May to July, and then rapidly developed into the La Niña event starting in August 2016 and continuing until December 2016.
Despite the rapid variability in climatic conditions, SQUARE-Mamba achieves strong predictive performance, primarily due to its TEB architecture, which successfully captures short-term variations.
Furthermore, in the rightmost red circle of Figure \ref{fig:moderate_2}, the SQUARE-Mamba achieves satisfactory prediction results even during the El Niño event in 2023 with short-term fluctuation, as mentioned before.
All in all, under the extreme conditions associated with the El Niño and La Niña events, SQUARE-Mamba consistently achieves higher DF accuracy than benchmark methods, further highlighting its reliability in DF, as demonstrated in the third and fourth rows of Table \ref{tab:quantitative}.

Finally, we evaluate the performance of the methods under the most challenging conditions for DF prediction, specifically dry climatic regions (i.e., Milparinka and Pooncarie), as presented in Figures \ref{fig:dry_1} and \ref{fig:dry_2}.
Even in extremely dry regions, SQUARE-Mamba consistently outperforms benchmark methods.
For example, as illustrated by the middle circle in Figure \ref{fig:dry_1}, SQUARE-Mamba has demonstrated robust and stable performance during the La Niña event of 2016 despite experiencing significant short-term fluctuations, as previously discussed.
Furthermore, the middle circle in Figure \ref{fig:dry_2} clearly shows that SQUARE-Mamba maintains outstanding prediction accuracy, even under severely dry conditions in 2019.
Notably, the first half of 2019 experienced the El Niño event, which then transitioned into a normal climatic phase lasting until August 2020.
However, despite this classification as a normal period (based on an ONI threshold of +/- 0.5$^\circ$C), Australia experienced the devastating and exceptionally severe wildfire season known as the Black Summer \cite{filkov2020impact}.
This highlights a critical issue: even seemingly climatically ``normal" conditions can result in severe economic losses due to unforeseen events.
Thus, providing accurate DF predictions is crucial for early detection and response.
As summarized in the last two rows of Table \ref{tab:quantitative}, SQUARE-Mamba consistently achieves superior quantitative results compared to benchmark models, underscoring its robustness and reliability even under the most demanding scenarios.

To sum up, we evaluated various scenarios across different climatic conditions, from wet to dry, over long-term periods (i.e., December 2007 to December 2023), including normal, El Ni\~{n}o, and La Ni\~{n}a years, to assess the robustness of the studied methods.
Table \ref{tab:quantitative} demonstrates that SQUARE-Mamba outperforms benchmark methods across all climatic regions, achieving an average improvement of more than $9.8\%$ in $\text{R}^2$.
Notably, SQUARE-Mamba maintains its superior performance even in the most challenging testing scenarios for DF (i.e., dry climatic regions), with $\text{R}^2$ exceeding $0.85$ in both Milparinka and Pooncarie.
Over the long term, Figures \ref{fig:wet_1} to \ref{fig:dry_2} illustrate the robustness and consistency of SQUARE-Mamba in adapting to climate variations compared to benchmark methods, which can be attributed to the original design philosophy behind the proposed temporal modeling (i.e., TEB in Figure \ref{fig:TEM}).

\subsection{Ablation Study}
\label{subsec:DFablation_study}
As demonstrated in Section \ref{subsec:exp}, SQUARE-Mamba achieves state-of-the-art performance in both qualitative and quantitative evaluations across various climatic regions.
To gain deeper insights into the efficacy of the proposed SEB and QLTEM, we conducted ablation studies, with the results summarized in Tables \ref{tab:ablation_wet}, \ref{tab:ablation_moderate}, and \ref{tab:ablation_dry}. 
Before discussing the details, we first briefly recall the design philosophy.
The proposed SEB integrates meteorological information from neighboring regions to capture spatial coherence, thereby enhancing prediction accuracy. 
Meanwhile, QLTEM is adopted to identify short-term climate fluctuations and is implemented using the quantum deep network (QUEEN) to capture more abstract temporal features through the quantum entanglement mechanism, providing informative features for the final prediction.
To sum up, we conducted ablation studies to evaluate the efficacy of the spatial analysis strategy (i.e., SEB) and the temporal analysis strategy (i.e., QLTEM) across three distinct climatic regions (i.e., wet, moderate, and dry).
In the following paragraph, we provide a detailed analysis of our findings, highlighting their impact on model performance in different climatic conditions.

As shown in the second and fourth rows of Tables \ref{tab:ablation_wet}, \ref{tab:ablation_moderate}, and \ref{tab:ablation_dry}, the integration of SEB does further improve DF accuracy across all performance metrics.
These observations clearly demonstrate the efficacy of SEB in enhancing DF predictions across different climatic conditions.
Furthermore, in the proposed TEB, QLTEM captures short-term climate variations by entangling temporal features.
As illustrated in the third and fourth rows of Tables \ref{tab:ablation_wet}, \ref{tab:ablation_moderate}, and \ref{tab:ablation_dry}, it is clear that DF performance is noticeably improved by incorporating QLTEM.
Based on the above findings, we conclude that integrating SEB or QLTEM can significantly enhance prediction accuracy.
Considering both the existing spatial coherence information and short-term fluctuation problem, we incorporate SEB and QLTEM simultaneously to capture more sophisticated interactions between spatial and temporal features, thereby extracting refined spatiotemporal representations.
The advantages are evident in the first and fourth rows of Tables \ref{tab:ablation_wet}, \ref{tab:ablation_moderate}, and \ref{tab:ablation_dry}, showing that integrating both SEB and QLTEM leads to substantial improvements in five out of six evaluated metrics.
Both QLTEM and SEB significantly enhance DF prediction accuracy across various climatic scenarios, ranging from wet to dry conditions, demonstrating their generalizability and effectiveness for DF prediction.
On average, SQUARE-Mamba improves over the Mamba baseline by more than 13.7\% in MAE, 11.9\% in RMSE, and 2.7\% in R$^2$, respectively.
These results clearly demonstrate the advantage of incorporating spatial information and quantum computing into our framework, particularly as SQURE-Mamba builds upon the naive (but already strong) baseline performance.
To sum up, SQUARE-Mamba achieves state-of-the-art performance and is thus a promising tool for forecasting drought conditions, potentially helping mitigate severe economic impacts.

%%===============Table Ablation in Wet Region=============%%
\begin{table}[t]
\caption{{Ablation study of the proposed SQUARE-Mamba algorithm in wet regions (i.e., Woombah and Geehi) to evaluate the efficacy of the spatial encoding block (SEB) and the quantum local temporal encoding module (QLTEM).}}
\renewcommand\arraystretch{1.15}
\begin{center}
\setlength{\tabcolsep}{1.5mm}
\scalebox{1}{
\begin{tabular}{c c c|c c c }
\hline
\hline
\makecell[c]{SEB} & \makecell[c]{QLTEM} & \makecell[c]{Location} & \makecell[c]{MAE ($\downarrow$)} & \makecell[c]{RMSE ($\downarrow$)} & \makecell[c]{$\text{R}^2 (\uparrow)$}

\\
\hline
\multirow{2}{*}{\ding{55}} & \multirow{2}{*}{\ding{55}} 
& Woombah & {\bf0.1608}  & 0.2300 & 0.9515
\\
&& Geehi & 0.1805  & 0.2414 & 0.9376
\\

\hline
\multirow{2}{*}{\ding{55}} & \multirow{2}{*}{\ding{51}} 
& Woombah & 0.1817  & 0.2623 & 0.9370
\\
&& Geehi & 0.1710  & 0.2277 & 0.9444
\\

\hline
\multirow{2}{*}{\ding{51}} & \multirow{2}{*}{\ding{55}}
& Woombah & 0.2132 & 0.2744 & 0.9310
\\
&& Geehi & 0.1944  & 0.2594 & 0.9279
\\

\hline
\multirow{2}{*}{\ding{51}} & \multirow{2}{*}{\ding{51}} 
& Woombah & 0.1663  & {\bf0.2250} & {\bf0.9536}
\\
&& Geehi & {\bf0.1553}  & {\bf0.2108} & {\bf0.9524}
\\

\hline
\hline
\end{tabular}
}
\label{tab:ablation_wet}
\end{center}
\end{table}

%%===============Table Ablation in Moderate Region=============%%
\begin{table}[t]
\caption{{Ablation study of the proposed SQUARE-Mamba algorithm in moderate regions (i.e., Enngonia and Jerilderie) to evaluate the efficacy of the spatial encoding block (SEB) and the quantum local temporal encoding module (QLTEM).}}
\renewcommand\arraystretch{1.15}
\begin{center}
\setlength{\tabcolsep}{1.5mm}
\scalebox{1}{
\begin{tabular}{c c c|c c c }
\hline
\hline
\makecell[c]{SEB} & \makecell[c]{QLTEM} & \makecell[c]{Location} & \makecell[c]{MAE ($\downarrow$)} & \makecell[c]{RMSE ($\downarrow$)} & \makecell[c]{$\text{R}^2 (\uparrow)$}

\\
\hline
\multirow{2}{*}{\ding{55}} & \multirow{2}{*}{\ding{55}} 
& Enngonia & 0.2254  & 0.3726 & 0.8095
\\
&& Jerilderie & 0.2013  & 0.3100 & 0.8709
\\

\hline
\multirow{2}{*}{\ding{55}} & \multirow{2}{*}{\ding{51}} 
& Enngonia & 0.2026  & 0.3005 & 0.8761
\\
&& Jerilderie & 0.2085  & 0.3098 & 0.8711
\\

\hline
\multirow{2}{*}{\ding{51}} & \multirow{2}{*}{\ding{55}}
& Enngonia & 0.2160  & 0.3048 & 0.8724
\\
&& Jerilderie & 0.2153  & 0.3183 & 0.8639
\\

\hline
\multirow{2}{*}{\ding{51}} & \multirow{2}{*}{\ding{51}} 
& Enngonia & {\bf0.1948}  & {\bf0.2867} & {\bf0.8872}
\\
&& Jerilderie & {\bf0.2007}  & {\bf0.2915} & {\bf0.8858}
\\

\hline
\hline
\end{tabular}
}
\label{tab:ablation_moderate}
\end{center}
\end{table}

%%===============Table Ablation in Dry Region=============%%
\begin{table}[t]
\caption{{Ablation study of the proposed SQUARE-Mamba algorithm in dry regions (i.e., Milparinka and Pooncarie) to evaluate the efficacy of the spatial encoding block (SEB) and the quantum local temporal encoding module (QLTEM).}}
\renewcommand\arraystretch{1.15}
\begin{center}
\setlength{\tabcolsep}{1.5mm}
\scalebox{1}{
\begin{tabular}{c c c|c c c }
\hline
\hline
\makecell[c]{SEB} & \makecell[c]{QLTEM} & \makecell[c]{Location} & \makecell[c]{MAE ($\downarrow$)} & \makecell[c]{RMSE ($\downarrow$)} & \makecell[c]{$\text{R}^2 (\uparrow)$}

\\
\hline
\multirow{2}{*}{\ding{55}} & \multirow{2}{*}{\ding{55}} 
& Milparinka & 0.2715  & 0.3943 & 0.7936
\\
&& Pooncarie & 0.2105  & 0.3139 & 0.9017
\\

\hline
\multirow{2}{*}{\ding{55}} & \multirow{2}{*}{\ding{51}} 
& Milparinka & 0.2627  & 0.3490 & 0.8383
\\
&& Pooncarie & 0.2188  & 0.3107 & 0.9037
\\

\hline
\multirow{2}{*}{\ding{51}} & \multirow{2}{*}{\ding{55}}
& Milparinka & 0.2831  & 0.3766 & 0.8117
\\
&& Pooncarie & 0.2078  & 0.2866 & 0.9180
\\

\hline
\multirow{2}{*}{\ding{51}} & \multirow{2}{*}{\ding{51}} 
& Milparinka & {\bf0.2407}  & {\bf0.3234} & {\bf0.8612}
\\
&& Pooncarie & {\bf0.1891}  & {\bf0.2677} & {\bf0.9285}
\\

\hline
\hline
\end{tabular}
}
\label{tab:ablation_dry}
\end{center}
\end{table}

%==============FLOPs Table==================%
\begin{table}[t]
% \scriptsize
\footnotesize 
\caption{Comparisons of the complexity across various deep learning-based DF methods. Here, ``M'' represents $10^6$.}

%\vspace{-0.3cm}
\renewcommand\arraystretch{1.2}
%\vspace{-0.1cm}
\begin{center}
\setlength{\tabcolsep}{1.5mm}
\scalebox{0.85}{
\begin{tabular}{c|c|c|c|c|c|c}
\hline
\hline

\multirow{1}{*}{Methods} & \multirow{1}{*}{LSTM} & \multirow{1}{*}{ConvLSTM} & \multirow{1}{*}{ViT} & \multirow{1}{*}{Informer} & \multirow{1}{*}{S-Mamba} & \multirow{1}{*}{SQUARE-Mamba}

%aligned
\\
\hline
FLOPs & 0.65M & 327.21M & 32.59M & 14.93M & 11.90M & \textbf{0.43M}
\\

\hline
\hline
\end{tabular}}
\label{tab:ALL_FLOPS}
\end{center}
% \vspace{-0.3cm}
\end{table}

%Shapley additive explanations (SHAP)
\subsection{Discussion, SHAP Analysis, and Future Challenge}\label{sec:Performance_Insights}
Motivated by information theory \cite{lin2024qmm,lin2024qhcd}, we propose a two-branch temporal encoder comprising LTEM and QLTEM, enabling explicit evaluation of whether unitary features from the QNN can improve the classical Mamba baseline.
Ablation studies (cf. the third and fourth rows in Tables \ref{tab:ablation_wet} to \ref{tab:ablation_dry}) confirm that integrating the QNN branch significantly improves model performance, thereby highlighting the efficacy of quantum unitary features even within the naive (but already strong) classical architecture.
Tables \ref{tab:quantitative} to \ref{tab:ablation_dry} further demonstrate that the Mamba baseline model generally surpasses LSTM- and Transformer-based approaches even without the quantum component.
For example, on the Geehi dataset, the classical Mamba model (cf. the third row of Table \ref{tab:ablation_wet}) achieves MAE, RMSE, and R$^2$ as (0.1944, 0.2594, and 0.9279), outperforming LSTM (0.4242, 0.5562, and 0.6689), ConvLSTM (0.3249, 0.4493, and 0.7839), ViT (0.3221, 0.4199, and 0.8113), Informer (0.3635, 0.5327, and 0.6962), and S-Mamba (0.4113, 0.5445, and 0.6827), as summarized in the second row of Table \ref{tab:quantitative}.
Moreover, on the Milparinka dataset, incorporating the QNN branch leads to improvements of 14.98\% in MAE, 14.13\% in RMSE, and 6.1\% in $\text{R}^2$ (cf. in the third and fourth rows of Table \ref{tab:ablation_dry}).
Beyond outperforming deep learning baselines, our method and other learning-based models consistently outperform the traditional non-AI statistical ARIMA approach.
Overall, the results across Tables \ref{tab:ablation_wet} to \ref{tab:ablation_dry} demonstrate that introducing the quantum module (i.e., QLTEM) further enhances the baseline's performance.
On average, SQUARE-Mamba improves over the Mamba baseline by more than 13.7\% in MAE, 11.9\% in RMSE, and 2.7\% in R$^2$.
This improvement stems from QNN’s ability to capture temporal quantum entanglement, effectively modeling short-term climatic fluctuations, which complements the classical Mamba module’s strength in long-term climate dynamics.
Thus, we believe this complementarity is a key contributor to the observed state-of-the-art performance.

To investigate the individual contributions of seven meteorological factors, we applied Shapley additive explanations (SHAP) analysis\footnote{\scriptsize\url{https://shap.readthedocs.io/en/latest/index.html}}, a widely adopted and interpretable AI technique.
The analysis was conducted using Pooncarie data, a representative dry climatic region.
The SHAP-derived contribution percentages for precipitation, potential evapotranspiration, maximum temperature, minimum temperature, mean temperature, vapor pressure, and cloud cover were 27.22\%, 23.89\%, 11.67\%, 11.11\%, 10\%, 8.89\%, and 7.22\%, respectively.
These results are consistent with traditional SPEI-based assessments, emphasizing the dominant roles of precipitation and evapotranspiration.
Moreover, the notable contributions of the three temperature-related variables further underscore their importance in DF.
These insights suggest that future research could prioritize the five most influential variables to enhance the development of data-driven DF algorithms.

As shown in Table \ref{tab:ALL_FLOPS}, the proposed SQUARE-Mamba achieves the highest computational efficiency, requiring only 0.43M FLOPs, which is significantly lower than all the other benchmark models.
This highlights its suitability for large-scale remote sensing applications.
Although current quantum hardware still faces noise-related challenges, we remain optimistic that future advances in fault-tolerant quantum computing will overcome these limitations.
Thus, noise was not explicitly addressed in our algorithm design; even so, the prediction accuracy of SQUARE-Mamba is quite high.
Despite the instability associated with quantum AI, it is important to note that the classical Mamba baseline already surpasses other LSTM- and Transformer-based models, demonstrating strong effectiveness for the DF task.
Moreover, another major challenge is the hardware limitation that may hinder the broader adoption of quantum AI in large-scale prediction-like tasks.
Specifically, quantum qubits are expected to remain limited, posing scalability challenges for continental- or global-scale weather forecasting.
In these scenarios, current quantum devices with around 100 low-error-rate qubits \cite{ibmheron} may be insufficient to model high-dimensional and complex time-series data effectively.
As highlighted in the HyperQUEEN paper \cite{hyperqueen}, developing effective classical deep compression techniques is essential for reducing high-dimensional meteorological data into compact and informative feature representations.
These compressed features can then be efficiently processed by near-term quantum computers for downstream forecasting tasks.
Therefore, effective information compression strategies will be critical to large-scale quantum forecasting models.

\section{Conclusion}\label{sec:conclusion}

We have leveraged the Mamba deep learning (a powerful temporal analysis modeling) and quantum deep network (QUEEN) to achieve state-of-the-art drought forecasting (DF) performance.
This integration is driven by two key factors.
First, quantum AI has shown strong potential in prediction-like tasks, making it suitable for meteorological applications like DF.
Second, our results demonstrate that incorporating QNNs does further improve the DF performance over the naive (but already strong) Mamba baseline.
These findings suggest that QNNs are theoretically robust, with quantum full expressibility guarantees, and practically effective for real-world prediction tasks.
The proposed DF method is termed SPEI-driven quantum spatially-aware Mamba network (SQUARE-Mamba), which yields a remarkable DF improvement of more than $9.8\%$ in the essential index R$^2$.
SQUARE-Mamba is generally applicable across diverse climate regions (including wet, dry, and moderate areas) and varying climate patterns (including El Niño, La Niña, and normal years) over a long testing period. 
As the ablation study demonstrated, the superiority of SQUARE-Mamba can be attributed to the proposed spatially-aware mechanism and the temporal quantum entanglement, where the former exploits the meteorological data from neighboring regions to assist the DF at a target region, while the latter employs QUEEN to capture the short-term climate variations for the subsequent long-term DF.
The proposed SQUARE-Mamba has successfully fused the seven extracted/refined spatiotemporal features corresponding to seven meteorological variables to accurately predict the DF results, well aligning with the observed SPEI drought conditions.
Therefore, SQUARE-Mamba is expected to serve as a reliable algorithm (i.e., Algorithm \ref{alg:SQUARE-Mamba}) for the monitoring and management of water resources amidst ongoing extreme climate changes.

\renewcommand{\thesubsection}{\Alph{subsection}}
\bibliography{ref_abb}

\begin{thebibliography}{10}
\providecommand{\url}[1]{#1}
\csname url@samestyle\endcsname
\providecommand{\newblock}{\relax}
\providecommand{\bibinfo}[2]{#2}
\providecommand{\BIBentrySTDinterwordspacing}{\spaceskip=0pt\relax}
\providecommand{\BIBentryALTinterwordstretchfactor}{4}
\providecommand{\BIBentryALTinterwordspacing}{\spaceskip=\fontdimen2\font plus
\BIBentryALTinterwordstretchfactor\fontdimen3\font minus
  \fontdimen4\font\relax}
\providecommand{\BIBforeignlanguage}[2]{{%
\expandafter\ifx\csname l@#1\endcsname\relax
\typeout{** WARNING: IEEEtran.bst: No hyphenation pattern has been}%
\typeout{** loaded for the language `#1'. Using the pattern for}%
\typeout{** the default language instead.}%
\else
\language=\csname l@#1\endcsname
\fi
#2}}
\providecommand{\BIBdecl}{\relax}
\BIBdecl

\bibitem{wilhite2016drought}
D.~A. Wilhite, \emph{Drought as A Natural Hazard: Concepts and
  Definitions}.\hskip 1em plus 0.5em minus 0.4em\relax Milton Park, Abingdon,
  Oxfordshire, UK: Routledge, 2016.

\bibitem{mao2015climate}
Y.~Mao, B.~Nijssen, and D.~P. Lettenmaier, ``Is climate change implicated in
  the 2013--2014 {C}alifornia drought? {A} hydrologic perspective,''
  \emph{Geophysical Research Letters}, vol.~42, no.~8, pp. 2805--2813, Mar.
  2015.

\bibitem{van2016drought}
A.~F. Van~Loon, T.~Gleeson, J.~Clark, A.~I. Van~Dijk, K.~Stahl, J.~Hannaford,
  G.~Di~Baldassarre, A.~J. Teuling, L.~M. Tallaksen, and R.~Uijlenhoet,
  ``Drought in the {A}nthropocene,'' \emph{Nature Geoscience}, vol.~9, no.~2,
  pp. 89--91, Feb. 2016.

\bibitem{ImmerzealScience2010}
W.~W. Immerzeel, L.~P. H.~V. Beek, and M.~F.~P. Bierkens, ``Climate change will
  affect the {A}sian water towers,'' \emph{Science}, vol. 328, no. 5984, pp.
  1382--1385, Jun. 2010.

\bibitem{taylor2013ground}
R.~G. Taylor, B.~Scanlon, P.~D{\"o}ll, M.~Rodell, R.~Van~Beek, Y.~Wada
  \emph{et~al.}, ``Ground water and climate change,'' \emph{Nature Climate
  Change}, vol.~3, no.~4, pp. 322--329, Nov. 2012.

\bibitem{florke2018water}
M.~Fl{\"o}rke, C.~Schneider, and R.~I. McDonald, ``Water competition between
  cities and agriculture driven by climate change and urban growth,''
  \emph{Nature Sustainability}, vol.~1, no.~1, pp. 51--58, Jan. 2018.

\bibitem{giovannini2016drought}
L.~Giovannini, X.~Ma, and A.~Huete, ``Drought resilience of {A}ustralian
  rangelands under intense hydroclimatic variability,'' in \emph{Proc. IEEE
  IGARSS}, Beijing, China, 10-15 Jul. 2016, pp. 5467--5469.

\bibitem{Xiang2023}
K.~Xiang, B.~Wang, D.~L. Liu, C.~Chen, C.~Waters, A.~Huete, and Q.~Yu,
  ``Probabilistic assessment of drought impacts on wheat yield in south-eastern
  {A}ustralia,'' \emph{Agricultural Water Management}, vol. 284, pp.
  108\,359--108\,359, Jun. 2023.

\bibitem{chang2016climate}
H.~Chang and M.~R. Bonnette, ``Climate change and water-related ecosystem
  services: {I}mpacts of drought in {C}alifornia, {USA},'' \emph{Ecosystem
  Health and Sustainability}, vol.~2, no.~12, pp. 1--19, Jun. 2017.

\bibitem{bond2008impacts}
N.~R. Bond, P.~S. Lake, and A.~H. Arthington, ``The impacts of drought on
  freshwater ecosystems: {A}n {A}ustralian perspective,'' \emph{Hydrobiologia},
  vol. 600, pp. 3--16, Feb. 2008.

\bibitem{zhang2019urban}
X.~Zhang, N.~Chen, H.~Sheng, C.~Ip, L.~Yang, Y.~Chen \emph{et~al.}, ``Urban
  drought challenge to 2030 sustainable development goals,'' \emph{Science of
  the Total Environment}, vol. 693, pp. 1--11, Nov. 2019.

\bibitem{salvia2021added}
M.~M. Salvia, N.~S{\'a}nchez, M.~Piles, R.~C. Ruscica,
  {\'A}.~Gonz{\'a}lez-Zamora, E.~Roitberg \emph{et~al.}, ``The added-value of
  remotely-sensed soil moisture data for agricultural drought detection in
  {A}rgentina,'' \emph{IEEE Journal of Selected Topics in Applied Earth
  Observations and Remote Sensing}, vol.~14, pp. 6487--6500, May 2021.

\bibitem{mckee1993relationship}
T.~B. McKee, N.~J. Doesken, and J.~Kleist, ``The relationship of drought
  frequency and duration to time scales,'' in \emph{Proc. Conf. Appl.
  Climatol.}, Anaheim, California, United States, 17-22 Jan. 1993, pp.
  179--183.

\bibitem{vicente2010multiscalar}
S.~M. Vicente-Serrano, S.~Beguer{\'\i}a, and J.~I. L{\'o}pez-Moreno, ``A
  multiscalar drought index sensitive to global warming: {T}he standardized
  precipitation evapotranspiration index,'' \emph{Journal of Climate}, vol.~23,
  no.~7, pp. 1696--1718, Apr. 2010.

\bibitem{vicente2012performance}
S.~M. Vicente-Serrano, S.~Beguer{\'\i}a, J.~Lorenzo-Lacruz, J.~J. Camarero,
  J.~I. L{\'o}pez-Moreno, C.~Azorin-Molina \emph{et~al.}, ``Performance of
  drought indices for ecological, agricultural, and hydrological
  applications,'' \emph{Earth Interactions}, vol.~16, no.~10, pp. 1--27, Sep.
  2012.

\bibitem{HUETE1988295}
A.~R. Huete, ``A soil-adjusted vegetation index {(SAVI)},'' \emph{Remote
  Sensing of Environment}, vol.~25, no.~3, pp. 295--309, Aug. 1988.

\bibitem{hochreiter1997long}
S.~Hochreiter and J.~Schmidhuber, ``Long short-term memory,'' \emph{Neural
  Computation}, vol.~9, no.~8, pp. 1735--1780, Nov. 1997.

\bibitem{zhou2021informer}
H.~Zhou, S.~Zhang, J.~Peng, S.~Zhang, J.~Li, H.~Xiong, and W.~Zhang,
  ``Informer: {B}eyond efficient transformer for long sequence time-series
  forecasting,'' in \emph{Proc. AAAI}, vol.~35, no.~12, Virtual Event, 2-9 Feb.
  2021, pp. 11\,106--11\,115.

\bibitem{box2015time}
G.~E. Box, G.~M. Jenkins, G.~C. Reinsel, and G.~M. Ljung, \emph{Time Series
  Analysis: Forecasting and Control}.\hskip 1em plus 0.5em minus 0.4em\relax
  Hoboken, New Jersey, USA: John Wiley \& Sons, 2015.

\bibitem{DIKSHIT2021111979}
A.~Dikshit, B.~Pradhan, and A.~Huete, ``An improved {SPEI} drought forecasting
  approach using the long short-term memory neural network,'' \emph{Journal of
  Environmental Management}, vol. 283, pp. 1--12, Apr. 2021.

\bibitem{medsker1999recurrent}
L.~Medsker and L.~C. Jain, \emph{Recurrent Neural Networks: Design and
  Applications}.\hskip 1em plus 0.5em minus 0.4em\relax Boca Raton, Florida,
  USA: CRC Press, 1999.

\bibitem{ConvLSTM}
X.~Shi, Z.~Chen, H.~Wang, D.-Y. Yeung, W.-K. Wong, and W.-C. Woo,
  ``Convolutional {LSTM} network: {A} machine learning approach for
  precipitation nowcasting,'' in \emph{Proc. NeurIPS}, Montréal, Canada, 7-12
  Sep. 2015, pp. 802--810.

\bibitem{chung2014empirical}
\BIBentryALTinterwordspacing
J.~Chung, C.~Gulcehre, K.~Cho, and Y.~Bengio, ``Empirical evaluation of gated
  recurrent neural networks on sequence modeling,'' \emph{arXiv preprint
  arXiv:1412.3555}, Dec. 2014. [Online]. Available:
  \url{https://arxiv.org/abs/1412.3555}
\BIBentrySTDinterwordspacing

\bibitem{cho2014learning}
\BIBentryALTinterwordspacing
K.~Cho, B.~Van~Merri{\"e}nboer, C.~Gulcehre, D.~Bahdanau, F.~Bougares,
  H.~Schwenk, and Y.~Bengio, ``Learning phrase representations using {RNN}
  encoder-decoder for statistical machine translation,'' \emph{arXiv preprint
  arXiv:1406.1078}, Jun. 2014. [Online]. Available:
  \url{https://arxiv.org/abs/1406.1078}
\BIBentrySTDinterwordspacing

\bibitem{shang2023application}
J.~Shang, B.~Zhao, H.~Hua, J.~Wei, G.~Qin, and G.~Chen, ``{A}pplication of
  {I}nformer model based on {SPEI} for drought forecasting,''
  \emph{Atmosphere}, vol.~14, no.~6, pp. 1--20, May 2023.

\bibitem{dong2022prediction}
H.~Dong, L.~Sun, and F.~Ouyang, ``Prediction of {PM}2.5 concentration based on
  {I}nformer,'' \emph{Environmental Engineering}, vol.~40, no.~6, pp. 48--54,
  Mar. 2022.

\bibitem{simplemamba}
Z.~Wang, F.~Kong, S.~Feng, M.~Wang, X.~Yang, H.~Zhao, D.~Wang, and Y.~Zhang,
  ``Is {M}amba effective for time series forecasting?'' \emph{Neurocomputing},
  vol. 619, pp. 1--14, Feb. 2025.

\bibitem{dosovitskiy2021an}
A.~Dosovitskiy, L.~Beyer, A.~Kolesnikov, D.~Weissenborn, X.~Zhai,
  T.~Unterthiner \emph{et~al.}, ``An image is worth 16x16 words: Transformers
  for image recognition at scale,'' in \emph{Proc. ICLR}, Virtual Event, 3-7
  May 2021.

\bibitem{hasan2023spi}
N.~A. Hasan, Y.~Dongkai, and F.~Al-Shibli, ``{SPI} and {SPEI} drought
  assessment and prediction using {TBATS} and {ARIMA} models, {J}ordan,''
  \emph{Water}, vol.~15, no.~20, pp. 1--32, Oct. 2023.

\bibitem{hyndman2018forecasting}
\BIBentryALTinterwordspacing
R.~Hyndman and G.~Athanasopoulos, \emph{Forecasting: Principles and
  Practice}.\hskip 1em plus 0.5em minus 0.4em\relax OTexts, 2018. [Online].
  Available: \url{https://books.google.com.tw/books?id=_bBhDwAAQBAJ}
\BIBentrySTDinterwordspacing

\bibitem{hyperqueen}
C.-H. Lin and Y.-Y. Chen, ``Hyper{QUEEN}: Hyperspectral quantum deep network
  for image restoration,'' \emph{IEEE Transactions on Geoscience and Remote
  Sensing}, vol.~61, pp. 1--20, May 2023.

\bibitem{lin2024qmm}
C.-H. Lin, P.-W. Tang, and A.~R. Huete, ``Quantum feature-empowered deep
  classification for fast mangrove mapping,'' \emph{IEEE Transactions on
  Geoscience and Remote Sensing}, vol.~63, pp. 1--13, Jan. 2025.

\bibitem{lin2024qhcd}
C.-H. Lin, T.-H. Lin, and J.~Chanussot, ``Quantum information-empowered graph
  neural network for hyperspectral change detection,'' \emph{IEEE Transactions
  on Geoscience and Remote Sensing}, vol.~62, pp. 1--15, Nov. 2024.

\bibitem{lin2024prime}
C.-H. Lin and J.-T. Lin, ``Prime: Blind multispectral unmixing using virtual
  quantum prism and convex geometry,'' \emph{IEEE Transactions on Geoscience
  and Remote Sensing}, vol.~63, pp. 1--16, Feb. 2025.

\bibitem{liu2021quantum}
N.~Liu, T.~Huang, J.~Gao, Z.~Xu, D.~Wang, and F.~Li, ``Quantum-enhanced deep
  learning-based lithology interpretation from well logs,'' \emph{IEEE
  Transactions on Geoscience and Remote Sensing}, vol.~60, pp. 1--13, Jun.
  2021.

\bibitem{pasetto2024kernel}
E.~Pasetto, M.~Riedel, K.~Michielsen, and G.~Cavallaro, ``Kernel approximation
  on a quantum annealer for remote sensing regression tasks,'' \emph{IEEE
  Journal of Selected Topics in Applied Earth Observations and Remote Sensing},
  vol.~17, pp. 3262--3269, Jan. 2024.

\bibitem{miroszewski2023detecting}
A.~Miroszewski, J.~Mielczarek, G.~Czelusta, F.~Szczepanek, B.~Grabowski,
  B.~Le~Saux, and J.~Nalepa, ``Detecting clouds in multispectral satellite
  images using quantum-kernel support vector machines,'' \emph{IEEE Journal of
  Selected Topics in Applied Earth Observations and Remote Sensing}, vol.~16,
  pp. 7601--7613, Aug. 2023.

\bibitem{fan2025hybrid}
F.~Fan, Y.~Shi, T.~Guggemos, and X.~X. Zhu, ``Hybrid quantum deep learning with
  superpixel encoding for {E}arth observation data classification,'' \emph{IEEE
  Transactions on Neural Networks and Learning Systems}, pp. 1--14, Jan. 2025.

\bibitem{sebastianelli2021}
A.~Sebastianelli, D.~A. Zaidenberg, D.~Spiller, B.~Le~Saux, and S.~L. Ullo,
  ``On circuit-based hybrid quantum neural networks for remote sensing imagery
  classification,'' \emph{IEEE Journal of Selected Topics in Applied Earth
  Observations and Remote Sensing}, vol.~15, pp. 565--580, Dec. 2021.

\bibitem{sebastianelli2025}
A.~Sebastianelli, F.~Mauro, G.~Ciabatti, D.~Spiller, B.~Le~Saux, P.~Gamba, and
  S.~Ullo, ``Quanv4{EO}: {E}mpowering {E}arth observation by means of
  quanvolutional neural networks,'' \emph{IEEE Transactions on Geoscience and
  Remote Sensing}, vol.~63, pp. 1--14, Mar. 2025.

\bibitem{henderson2020}
M.~Henderson, S.~Shakya, S.~Pradhan, and T.~Cook, ``Quanvolutional neural
  networks: {P}owering image recognition with quantum circuits,'' \emph{Quantum
  Machine Intelligence}, vol.~2, no.~1, pp. 1--7, Feb. 2020.

\bibitem{hyperking}
C.-H. Lin and S.-S. Young, ``Hyper{KING}: Quantum-classical generative
  adversarial networks for hyperspectral image restoration,'' \emph{IEEE
  Transactions on Geoscience and Remote Sensing}, pp. 1--19, Apr. 2025.

\bibitem{otgonbaatar2023}
S.~Otgonbaatar and D.~Kranzlm{\"u}ller, ``Exploiting the quantum advantage for
  satellite image processing: {R}eview and assessment,'' \emph{IEEE
  Transactions on Quantum Engineering}, vol.~5, pp. 1--9, Dec. 2023.

\bibitem{Hong2024qsif}
Y.-Y. Hong, D.~Josh Domingo~Lopez, and Y.-Y. Wang, ``Solar irradiance
  forecasting using a hybrid quantum neural network: {A} comparison on
  {GPU}-based workflow development platforms,'' \emph{IEEE Access}, vol.~12,
  pp. 145\,079--145\,094, Oct. 2024.

\bibitem{PAQUET2022116583}
E.~Paquet and F.~Soleymani, ``Quantum{L}eap: {H}ybrid quantum neural network
  for financial predictions,'' \emph{Expert Systems with Applications}, vol.
  195, pp. 1--10, Jun. 2022.

\bibitem{Safari2021qwf}
A.~Safari and A.~A. Ghavifekr, ``Quantum neural networks {(QNN)} application in
  weather prediction of smart grids,'' in \emph{Proc. IEEE SGC}, Tabriz, Iran,
  7-9 Dec. 2021, pp. 1--6.

\bibitem{qu2022temporal}
Z.~Qu, X.~Liu, and M.~Zheng, ``Temporal-spatial quantum graph convolutional
  neural network based on {S}chr\"{o}dinger approach for traffic congestion
  prediction,'' \emph{IEEE Transactions on Intelligent Transportation Systems},
  vol.~24, no.~8, pp. 8677--8686, Sep. 2022.

\bibitem{gupta2024multiple}
I.~Gupta, D.~Saxena, A.~K. Singh, and C.-N. Lee, ``A multiple controlled
  {T}offoli driven adaptive quantum neural network model for dynamic workload
  prediction in cloud environments,'' \emph{IEEE Transactions on Pattern
  Analysis and Machine Intelligence}, vol.~46, pp. 1--15, May 2024.

\bibitem{habibi2025electrical}
M.~R. Habibi, S.~Golestan, Y.~Wu, J.~M. Guerrero, and J.~C. Vasquez,
  ``Electrical load forecasting in power systems based on quantum computing
  using time series-based quantum artificial intelligence,'' \emph{Scientific
  Reports}, vol.~15, no.~1, pp. 1--26, Mar. 2025.

\bibitem{fan2023hybrid}
F.~Fan, Y.~Shi, T.~Guggemos, and X.~X. Zhu, ``Hybrid quantum-classical
  convolutional neural network model for image classification,'' \emph{IEEE
  Transactions on Neural Networks and Learning Systems}, vol.~35, pp.
  18\,145--18\,159, Sep. 2023.

\bibitem{gu2023mamba}
\BIBentryALTinterwordspacing
A.~Gu and T.~Dao, ``Mamba: {L}inear-time sequence modeling with selective state
  spaces,'' \emph{arXiv preprint arXiv:2312.00752}, Dec. 2023. [Online].
  Available: \url{https://arxiv.org/abs/2312.00752}
\BIBentrySTDinterwordspacing

\bibitem{liang2024bi}
\BIBentryALTinterwordspacing
A.~Liang, X.~Jiang, Y.~Sun, and C.~Lu, ``{B}i-{M}amba4{TS}: {B}idirectional
  {M}amba for time series forecasting,'' \emph{arXiv preprint
  arXiv:2404.15772}, Apr. 2024. [Online]. Available:
  \url{https://arxiv.org/abs/2404.15772}
\BIBentrySTDinterwordspacing

\bibitem{yue2024biomamba}
\BIBentryALTinterwordspacing
L.~Yue, S.~Xing, Y.~Lu, and T.~Fu, ``Bio{M}amba: {A} pre-trained biomedical
  language representation model leveraging mamba,'' \emph{arXiv preprint
  arXiv:2408.02600}, Aug. 2024. [Online]. Available:
  \url{https://arxiv.org/abs/2408.02600}
\BIBentrySTDinterwordspacing

\bibitem{Li2024mambahsi}
Y.~Li, Y.~Luo, L.~Zhang, Z.~Wang, and B.~Du, ``Mamba{HSI}: {S}patial–spectral
  {M}amba for hyperspectral image classification,'' \emph{IEEE Transactions on
  Geoscience and Remote Sensing}, vol.~62, pp. 1--16, Jul. 2024.

\bibitem{vaswani2017attention}
A.~Vaswani, N.~Shazeer, N.~Parmar, J.~Uszkoreit, L.~Jones, A.~N. Gomez,
  {\L}.~Kaiser, and I.~Polosukhin, ``Attention is all you need,'' in
  \emph{Proc. NeurIPS}, Long Beach, California, USA, 4-9 Dec. 2017, pp.
  6000--6010.

\bibitem{tang2024transformer}
P.-W. Tang, C.-H. Lin, and Y.~Liu, ``Transformer-driven inverse problem
  transform for fast blind hyperspectral image dehazing,'' \emph{IEEE
  Transactions on Geoscience and Remote Sensing}, vol.~62, pp. 1--14, Jan.
  2024.

\bibitem{Young2024tnnls}
S.-S. Young, C.-H. Lin, and Z.-C. Leng, ``Unsupervised abundance matrix
  reconstruction transformer-guided fractional attention mechanism for
  hyperspectral anomaly detection,'' \emph{IEEE Transactions on Neural Networks
  and Learning Systems}, pp. 1--15, Aug. 2024.

\bibitem{Gnanasambandamtanhpami2023}
R.~Gnanasambandam, B.~Shen, J.~Chung, X.~Yue, and Z.~Kong, ``Self-scalable tanh
  ({S}tan): {M}ulti-scale solutions for physics-informed neural networks,''
  \emph{IEEE Transactions on Pattern Analysis and Machine Intelligence},
  vol.~45, no.~12, pp. 15\,588--15\,603, Dec. 2023.

\bibitem{nearestpadding}
U.~Pujianto, A.~P. Wibawa, M.~I. Akbar, and D.~M.~P. Murti, ``K-nearest
  neighbor {(K-NN)} based missing data imputation,'' in \emph{Proc. IEEE
  ICSITech}, Yogyakarta, Indonesia, 23--24 Oct. 2019, pp. 83--88.

\bibitem{lin2024superrpca}
J.-T. Lin and C.-H. Lin, ``Super{RPCA}: {A} collaborative superpixel
  representation prior-aided {RPCA} for hyperspectral anomaly detection,''
  \emph{IEEE Transactions on Geoscience and Remote Sensing}, vol.~62, pp.
  1--16, Sep. 2024.

\bibitem{maas2013rectifier}
A.~L. Maas, A.~Y. Hannun, and A.~Y. Ng, ``Rectifier nonlinearities improve
  neural network acoustic models,'' in \emph{Proc. ICML}, vol.~30, no.~1.\hskip
  1em plus 0.5em minus 0.4em\relax Atlanta, GA, 16-21 Jun. 2013, pp. 1--6.

\bibitem{extremeclimate}
D.~R. Easterling, J.~L. Evans, P.~Y. Groisman, T.~R. Karl, K.~E. Kunkel, and
  P.~Ambenje, ``Observed variability and trends in extreme climate events: {A}
  brief review,'' \emph{Bulletin of the American Meteorological Society},
  vol.~81, no.~3, pp. 417--426, Mar. 2000.

\bibitem{han2024demystify}
D.~Han, Z.~Wang, Z.~Xia, Y.~Han, Y.~Pu, C.~Ge, J.~Song, S.~Song, B.~Zheng, and
  G.~Huang, ``Demystify {M}amba in vision: {A} linear attention perspective,''
  in \emph{Proc. NeurIPS}, Vancouver, Canada, 10-15 Dec. 2024, pp. 1--16.

\bibitem{silu}
B.~Singh, S.~Patel, A.~Vijayvargiya, and R.~Kumar, ``Analyzing the impact of
  activation functions on the performance of the data-driven gait model,''
  \emph{Results in Engineering}, vol.~18, pp. 1--13, Jun. 2023.

\bibitem{elu}
L.~Trottier, P.~Giguere, and B.~Chaib-Draa, ``Parametric exponential linear
  unit for deep convolutional neural networks,'' in \emph{Proc. IEEE ICMLA},
  Cancun, Mexico, 18-21 Dec. 2017, pp. 207--214.

\bibitem{lin2024quantum}
C.-H. Lin, C.-Y. Kuo, and S.-S. Young, ``Quantum adversarial learning for
  hyperspectral remote sensing,'' in \emph{Proc. IEEE IGARSS}, Athens, Greece,
  7-12 Jul. 2024, pp. 7807--7811.

\bibitem{lin2023quantum}
C.-H. Lin and Y.-Y. Chen, ``Quantum deep hyperspectral satellite remote
  sensing,'' in \emph{IEEE IGARSS}, Pasadena, California, 16-21 Jul. 2023, pp.
  7316--7319.

\bibitem{Qunitary}
M.~A. Nielsen and I.~L. Chuang, \emph{Quantum Computation and Quantum
  Information}.\hskip 1em plus 0.5em minus 0.4em\relax Shaftesbury Road,
  Cambridge, United Kingdom: Cambridge University Press, Cambridge, 2010.

\bibitem{dropout}
\BIBentryALTinterwordspacing
G.~E. Hinton, N.~Srivastava, A.~Krizhevsky, I.~Sutskever, and R.~R.
  Salakhutdinov, ``Improving neural networks by preventing co-adaptation of
  feature detectors,'' \emph{arXiv preprint arXiv:1207.0580}, Jul. 2012.
  [Online]. Available: \url{https://arxiv.org/abs/1207.0580}
\BIBentrySTDinterwordspacing

\bibitem{gelu}
\BIBentryALTinterwordspacing
D.~Hendrycks and K.~Gimpel, ``Gaussian error linear units ({GELUs}),''
  \emph{arXiv preprint arXiv:1606.08415}, Jun. 2016. [Online]. Available:
  \url{https://arxiv.org/abs/1606.08415}
\BIBentrySTDinterwordspacing

\bibitem{harris2020version}
I.~Harris, T.~J. Osborn, P.~Jones, and D.~Lister, ``Version 4 of the {CRU} {TS}
  monthly high-resolution gridded multivariate climate dataset,''
  \emph{Scientific Data}, vol.~7, no.~1, pp. 1--18, Mar. 2020.

\bibitem{nagavciuc2019stable}
V.~Nagavciuc, M.~Ionita, A.~Perșoiu, I.~Popa, N.~J. Loader, and D.~McCarroll,
  ``Stable oxygen isotopes in {R}omanian oak tree rings record summer droughts
  and associated large-scale circulation patterns over {E}urope,''
  \emph{Climate Dynamics}, vol.~52, pp. 6557--6568, Nov. 2019.

\bibitem{renard2019national}
D.~Renard and D.~Tilman, ``National food production stabilized by crop
  diversity,'' \emph{Nature}, vol. 571, no. 7764, pp. 257--260, May 2019.

\bibitem{rhee2017meteorological}
J.~Rhee and J.~Im, ``{M}eteorological drought forecasting for ungauged areas
  based on machine learning: {U}sing long-range climate forecast and remote
  sensing data,'' \emph{Agricultural and Forest Meteorology}, vol. 237, pp.
  105--122, May 2017.

\bibitem{belayneh2014long}
A.~Belayneh, J.~Adamowski, B.~Khalil, and B.~Ozga-Zielinski, ``Long-term {SPI}
  drought forecasting in the {A}wash {R}iver {B}asin in {E}thiopia using
  wavelet neural network and wavelet support vector regression models,''
  \emph{Journal of Hydrology}, vol. 508, pp. 418--429, Jan. 2014.

\bibitem{codemm}
C.-H. Lin, M.-C. Chu, and P.-W. Tang, ``{CODE-MM}: Convex deep mangrove mapping
  algorithm based on optical satellite images,'' \emph{IEEE Transactions on
  Geoscience and Remote Sensing}, vol.~61, pp. 1--19, Sep. 2023.

\bibitem{zhang2005neural}
G.~P. Zhang and M.~Qi, ``Neural network forecasting for seasonal and trend time
  series,'' \emph{European Journal of Operational Research}, vol. 160, no.~2,
  pp. 501--514, Jan. 2005.

\bibitem{wang2024deep}
\BIBentryALTinterwordspacing
Y.~Wang, H.~Wu, J.~Dong, Y.~Liu, M.~Long, and J.~Wang, ``Deep time series
  models: {A} comprehensive survey and benchmark,'' \emph{arXiv preprint
  arXiv:2407.13278}, Jul. 2024. [Online]. Available:
  \url{https://arxiv.org/abs/2407.13278}
\BIBentrySTDinterwordspacing

\bibitem{loshchilov2018decoupled}
I.~Loshchilov and F.~Hutter, ``Decoupled weight decay regularization,'' in
  \emph{Proc. ICLR}, New Orleans, USA, 6-9 May 2019, pp. 1--8.

\bibitem{pennylane}
\BIBentryALTinterwordspacing
V.~Bergholm, J.~Izaac, M.~Schuld, C.~Gogolin, S.~Ahmed, V.~Ajith \emph{et~al.},
  ``Pennylane: Automatic differentiation of hybrid quantum-classical
  computations,'' \emph{arXiv preprint arXiv:1811.04968}, Nov. 2018. [Online].
  Available: \url{https://arxiv.org/abs/1811.04968}
\BIBentrySTDinterwordspacing

\bibitem{pandey2024deepfake}
A.~Pandey and B.~Rudra, ``Deepfake audio detection using quantum learning
  models,'' in \emph{Proc. IEEE MECOM}, Abu Dhabi, United Arab Emirates, 17-20
  Nov. 2024, pp. 1--6.

\bibitem{yu2024application}
W.~Yu, L.~Yin, C.~Zhang, Y.~Chen, and A.~X. Liu, ``Application of quantum
  recurrent neural network in low resource language text classification,''
  \emph{IEEE Transactions on Quantum Engineering}, vol.~5, pp. 1--13, Mar.
  2024.

\bibitem{alvarez2018quantum}
U.~Alvarez-Rodriguez, M.~Sanz, L.~Lamata, and E.~Solano, ``Quantum artificial
  life in an {IBM} quantum computer,'' \emph{Scientific Reports}, vol.~8,
  no.~1, pp. 1--9, Oct. 2018.

\bibitem{haghparast2024innovative}
M.~Haghparast, E.~Moguel, J.~Garcia-Alonso, T.~Mikkonen, and J.~M. Murillo,
  ``Innovative approaches to teaching quantum computer programming and quantum
  software engineering,'' in \emph{Proc. IEEE QCE}, vol.~2, Montr\'eal,
  Qu\'ebec, Canada, 15-20 Sep. 2024, pp. 251--255.

\bibitem{dikshit2020short}
A.~Dikshit, B.~Pradhan, and A.~M. Alamri, ``Short-term spatio-temporal drought
  forecasting using random forests model at {New South Wales, Australia},''
  \emph{Applied Sciences}, vol.~10, no.~12, pp. 1--16, Jun. 2020.

\bibitem{willmott2005advantages}
C.~J. Willmott and K.~Matsuura, ``Advantages of the mean absolute error {(MAE)}
  over the root mean square error {(RMSE)} in assessing average model
  performance,'' \emph{Climate Research}, vol.~30, pp. 79--82, Dec. 2005.

\bibitem{chai2014root}
T.~Chai and R.~R. Draxler, ``Root mean square error {(RMSE)} or mean absolute
  error {(MAE)},'' \emph{Geoscientific Model Development Discussions}, vol.~7,
  no.~1, pp. 1525--1534, Feb. 2014.

\bibitem{cameron1997r}
A.~C. Cameron and F.~A. Windmeijer, ``An {R}-squared measure of goodness of fit
  for some common nonlinear regression models,'' \emph{Journal of
  Econometrics}, vol.~77, no.~2, pp. 329--342, Apr. 1997.

\bibitem{oni}
M.~H. Glantz and I.~J. Ramirez, ``Reviewing the {O}ceanic {N}i{\~n}o {I}ndex
  ({ONI}) to enhance societal readiness for {El Ni{\~n}o’s impacts},''
  \emph{International Journal of Disaster Risk Science}, vol.~11, pp. 394--403,
  May 2020.

\bibitem{filkov2020impact}
A.~I. Filkov, T.~Ngo, S.~Matthews, S.~Telfer, and T.~D. Penman, ``Impact of
  {A}ustralia's catastrophic 2019/20 bushfire season on communities and
  environment. retrospective analysis and current trends,'' \emph{Journal of
  Safety Science and Resilience}, vol.~1, no.~1, pp. 44--56, Sep. 2020.

\bibitem{ibmheron}
D.~Castelvecchi, ``{IBM} releases first-ever 1,000-qubit quantum chip,''
  \emph{Nature}, vol. 624, no. 7991, pp. 238--238, Dec. 2023.

\end{thebibliography}

\begin{IEEEbiography}[{\resizebox{0.95in}{!}{\includegraphics[width=1in,height=1.25in,clip,keepaspectratio]{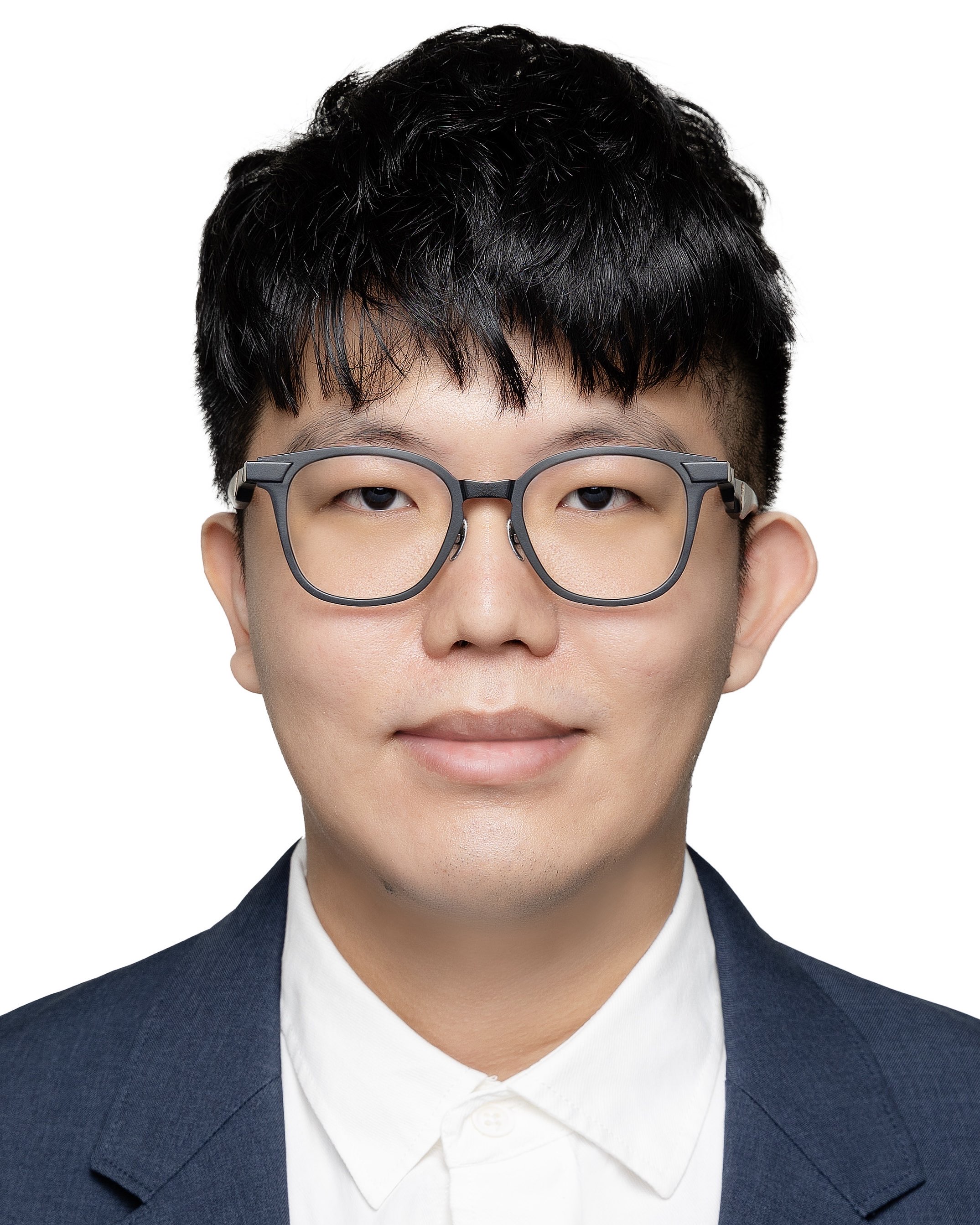}}}]
{\bf Po-Wei Tang}
(S'20)
received the B.S. degree from the Department of Electronic Engineering, National Changhua University of Education, Changhua, Taiwan, in 2018.
He is currently pursuing the Ph.D. degree with the Intelligent Hyperspectral Computing Laboratory, Institute of Computer and Communication Engineering, National Cheng Kung University (NCKU), Tainan, Taiwan.
His research interests include deep learning, convex optimization, tensor completion, and hyperspectral imaging.

Mr. Tang has received a highly competitive scholarship from NCKU, as well as the Pan Wen Yuan Award from the Industrial Technology Research Institute (ITRI) of Taiwan.
He has been selected as a recipient for the Ph.D. Students Study Abroad Program from the National Science and Technology Council (NSTC), in 2024.
\end{IEEEbiography}
 
\begin{IEEEbiography}[{\resizebox{0.95in}{!}{\includegraphics[width=1in,height=1.25in,clip,keepaspectratio]{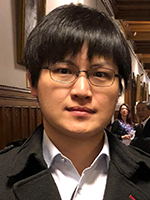}}}]
{\bf Chia-Hsiang Lin}
(S'10-M'18-SM'24)
received the B.S. degree in electrical engineering and the Ph.D. degree in communications engineering from National Tsing Hua University (NTHU), Taiwan, in 2010 and 2016, respectively.
From 2015 to 2016, he was a Visiting Student of Virginia Tech,
Arlington, VA, USA.

He is currently an Associate Professor with the Department of Electrical Engineering, and also with 
the Miin Wu School of Computing,
National Cheng Kung University (NCKU), Taiwan.
Before joining NCKU, he held research positions with The Chinese University of Hong Kong, HK (2014 and 2017), 
NTHU (2016-2017), 
and the University of Lisbon (ULisboa), Lisbon, Portugal (2017-2018).
He was an Assistant Professor with the Center for Space and Remote Sensing Research, National Central University, Taiwan, in 2018, and a Visiting Professor with ULisboa, in 2019.
His research interests include network science, 
quantum computing,
convex geometry and optimization, blind signal processing, and imaging science.

Dr. Lin received the Emerging Young Scholar Award (The 2030 Cross-Generation Program) from National Science and Technology Council (NSTC), from 2023 to 2027,
the Future Technology Award from NSTC, in 2022,
the Outstanding Youth Electrical Engineer Award from The Chinese Institute of Electrical Engineering (CIEE), in 2022,
the Best Young Professional Member Award from IEEE Tainan Section, in 2021,
the Prize Paper Award from IEEE Geoscience and Remote Sensing Society (GRS-S), in 2020, 
the Top Performance Award from Social Media Prediction Challenge at ACM Multimedia, in 2020,
and The 3rd Place from AIM Real World Super-Resolution Challenge at IEEE International Conference on Computer Vision (ICCV), in 2019. 
He received the Ministry of Science and Technology (MOST) Young Scholar Fellowship, together with the EINSTEIN Grant Award, from 2018 to 2023.
In 2016, he was a recipient of the Outstanding Doctoral Dissertation Award from the Chinese Image Processing and Pattern Recognition Society and the Best Doctoral Dissertation Award from the IEEE GRS-S.
\end{IEEEbiography}

\begin{IEEEbiography}[{\resizebox{0.97in}{!}{\includegraphics[width=1in,height=1.25in,clip,keepaspectratio]{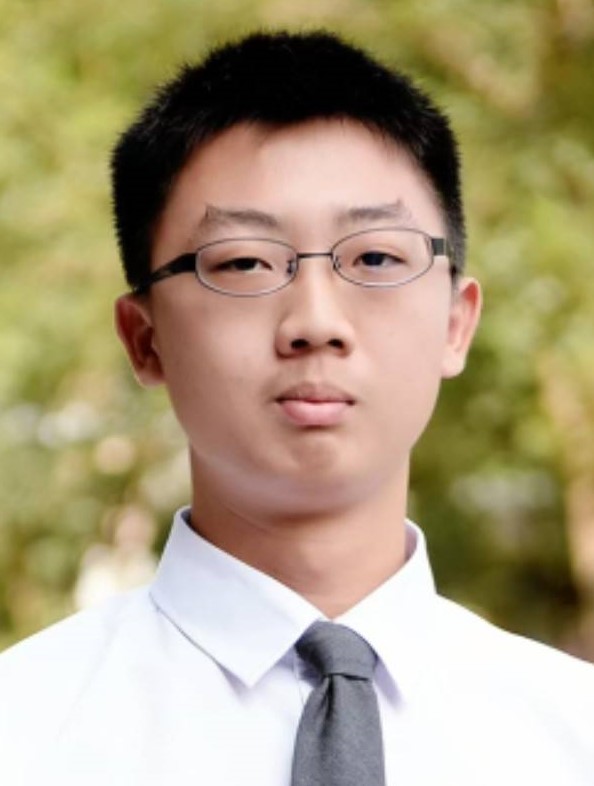}}}]
{\bf Jian-Kai Huang} (S'25) received the B.S. degree from the Department of Electrical Engineering, National Chung Cheng University, Chiayi, Taiwan, in 2023. 
He is currently pursuing the Ph.D. degree with the Intelligent Hyperspectral Computing Laboratory, Institute of Computer and Communication Engineering, National Cheng Kung University (NCKU), Tainan, Taiwan. 
His research interests include deep learning, drought forecasting, and hyperspectral imaging. 

\end{IEEEbiography}

\begin{IEEEbiography}[{\resizebox{0.97in}{!}{\includegraphics[width=1in,height=1.25in,clip,keepaspectratio]{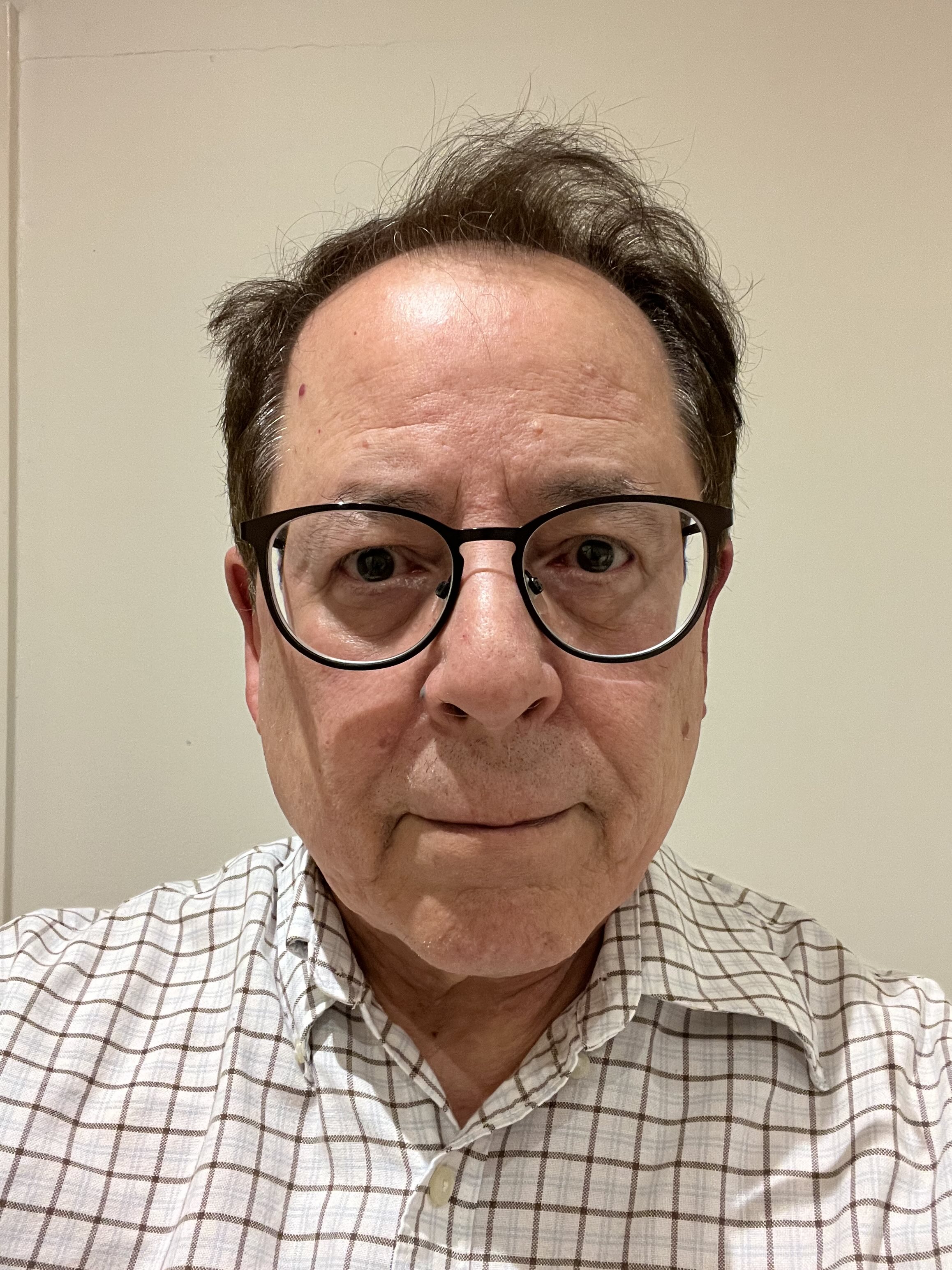}}}]
{\bf Alfredo R. Huete} received the M.Sc. degree in soil and plant biology from the University of California at Berkeley, Berkeley, CA, USA, in 1982, and the Ph.D. degree in soil and water science from The University of Arizona, Tucson, AZ, USA, in 1984. From 1984 to 2009, he was an Assistant Professor and an Associate Professor with The University of Arizona. He has over 20 years of experience in working on satellite mission teams, including the NASA-EOS MODIS Science Team, the New Millennium EO-1 Hyperion Team, and JAXA GLI Team, through which he developed the soil-adjusted vegetation index (SAVI) and the enhanced vegetation index (EVI). He is currently a Distinguished Professor with the University of Technology Sydney, Sydney, NSW, Australia, where he leads the Ecosystem Dynamics Health and Resilience Research Program in the Faculty of Science and is a Core Member of the Centre for Advanced Modelling and Geospatial Information Systems, Faculty of Engineering and Information Technology. He also leads the development of vegetation products for the Australian Terrestrial Ecosystem Research Network (TERN).
\end{IEEEbiography}

\end{document}